\documentclass[aps,prd,twocolumn,showpacs,floatfix,preprintnumbers,amsfont,amsmath,amssymb,nofootinbib, superscriptaddress]{revtex4-2}
\usepackage{graphicx} 
\usepackage{amsmath,amssymb}

\usepackage{aas_macros} % to let the compiler know what the abbreviations for the journals are (in the .bib file). 

\usepackage{lineno,hyperref}
\hypersetup{
 colorlinks=true,
 linktoc=all,
 linkcolor=blue,
 citecolor=blue,
 urlcolor = blue}

\usepackage{soul}
\usepackage[normalem]{ulem} 
\usepackage[dvipsnames]{xcolor}

\usepackage{orcidlink}

\def\au{{\rm AU}}
\def\pc{{\rm pc}}
\def\kms{{\rm km/s}}
\def\msun{{\rm M}_\odot}

\def\ymax{y_{\rm max}}
\def\yGO{y_{\rm GO}}
\def\snr{{\rm SNR}}

\def\dLS{d_{\rm LS}}
\def\ML{m_{\rm L}}
\def\MS{m_{\rm S}}

\def\rs{R_{\rm L}} 
\def\re{R_{\rm E}}

\begin{document}

\title{Constraining the environment of compact binary mergers with self-lensing signatures}

\author{Helena Ubach\,\orcidlink{0000-0002-0679-9074}}
\email{helenaubach@icc.ub.edu}
\affiliation{Departament de F\'{i}sica Qu\`{a}ntica i Astrof\'{i}sica (FQA), Universitat de Barcelona (UB),  c. Mart\'{i} i Franqu\`{e}s, 1, 08028 Barcelona, Spain}
\affiliation{Institut de Ci\`{e}ncies del Cosmos (ICCUB), Universitat de Barcelona (UB), c. Mart\'{i} i Franqu\`{e}s, 1, 08028 Barcelona, Spain}
\author{Mark Gieles\,\orcidlink{0000-0002-9716-1868}}
\affiliation{Institut de Ci\`{e}ncies del Cosmos (ICCUB), Universitat de Barcelona (UB), c. Mart\'{i} i Franqu\`{e}s, 1, 08028 Barcelona, Spain}
\affiliation{ICREA, Pg. Llu\'{i}s Companys 23, 08010 Barcelona, Spain}
\affiliation{Institut d'Estudis Espacials de Catalunya (IEEC), Edifici RDIT, Campus UPC, 08860 Castelldefels (Barcelona), Spain}\author{Jordi Miralda-Escud\'{e}\,\orcidlink{0000-0002-2316-8370}}
\affiliation{Departament de F\'{i}sica Qu\`{a}ntica i Astrof\'{i}sica (FQA), Universitat de Barcelona (UB),  c. Mart\'{i} i Franqu\`{e}s, 1, 08028 Barcelona, Spain}
\affiliation{Institut de Ci\`{e}ncies del Cosmos (ICCUB), Universitat de Barcelona (UB), c. Mart\'{i} i Franqu\`{e}s, 1, 08028 Barcelona, Spain}
\affiliation{ICREA, Pg. Llu\'{i}s Companys 23, 08010 Barcelona, Spain}\affiliation{Institut d'Estudis Espacials de Catalunya (IEEC), Edifici RDIT, Campus UPC, 08860 Castelldefels (Barcelona), Spain}

\begin{abstract}
Gravitational waves (GWs) from coalescing binary black holes 
 (BBHs) can come from different environments. 
GWs interact gravitationally with astrophysical objects, which makes it possible 
to use gravitational lensing by objects in the same gravitational system (self-lensing) to learn about their environments. 
We quantify the probability of self-lensing through the optical depth $\tau$ for the main channels of detectable GWs at frequencies $f_{\rm GW}\sim (1-10^3)\,{\rm Hz}$.  We then analyze the detectability of the lensing effect (imprint). 
In star clusters, the probability of self-lensing by stellar-mass black holes (BHs) is low, $\tau\simeq10^{-7}$, even when taking into account nearby BHs in resonant interactions, $\tau\simeq 10^{-5}$.  
Additionally, the lensing imprint of a stellar-mass lens (diffraction and interference) is too marginal to be detectable by the LIGO-Virgo-KAGRA detectors and most Einstein Telescope signals.
For a massive BH lens 
in the center of a cluster, the probability can reach $\tau\simeq 10^{-4}$ either via von Zeipel-Lidov-Kozai induced mergers of BBHs orbiting a central massive BH, or BBHs formed as GW captures in single-single interactions in the Bahcall-Wolf cusp of a nuclear cluster. For self-lensing by a supermassive BH for BBHs in the migration trap of an active galactic nucleus (AGN) disk, $\tau \simeq 10^{-2}$. 
The imprint of these massive lenses are multiple images that are already detectable. Moreover, self-lensed signals from AGN disks have a distinct linear polarization. 
The probability depends on the extent of the detectability through the threshold impact parameter $\ymax$, which can increase for future detectors. 
We conclude that constraining the environment of BBHs is possible by combining self-lensing imprints with other waveform signatures such as eccentricity and polarization. 
\end{abstract}

\maketitle

\section{Introduction}

Gravitational waves (GWs) are oscillations of the space-time metric that are created by accelerating masses \cite{einstein-gw-16}.
Nearly 300 GW events from mergers of binary compact objects have been confirmed by the LIGO-Virgo-KAGRA (LVK) detectors \cite{LIGO18-GWTC1,LIGO20-GWTC2,LIGO21-GWTC3,venumadhav-20,zackay-21,LIGO25-GWTC4}. A crucial open question remains what the origin of these mergers is. Binary black hole (BBH) mergers can form either in isolation as the result of the evolution of a massive stellar binary \cite{belczynski-02,mandel-broekgaarden-22}, or dynamically at the center of dense stellar systems, such as open clusters \cite{banerjee-17,rastello-19,kumamoto-20}, globular clusters (GCs) \cite{portegies-00,rodriguez-16,antonini-23},  nuclear clusters (NCs) with \cite{oleary-09,hoang-18} or without \cite{antonini-rasio-16} a massive central black hole (BH) and active galactic nuclei (AGNs) \cite{mckernan-12,bellovary-16,stone-17,bartos-17,mckernan-18}. 

If the BBH interacts gravitationally with other massive objects, the GW signal can reveal information about the environment of the merging binaries,
for example via Doppler phase shifts \cite{samsing-24a,hendriks-24,samsing-24b} or variable gravitational redshifts acquired in the emission region. GWs can be further altered by gravitational lensing during propagation when they
are deflected by intervening masses \cite{einstein-16-deflection}.
Although gravitational lensing is usually thought to occur far from the source by unrelated foreground galaxies \cite[e.g.,][]{schneider-92},
it can also occur close to the source and reveal information about the BBH environment. 

In this paper we study the environmental signatures 
caused by {\it self-lensing}, which we define as gravitational lensing caused by masses in the 
same gravitationally bound system as the source, 
similar to \cite{gould-disk-95}\footnote{The term is sometimes used in the more restricted case of lensing of light in binary systems \cite{gould-binary-95,beskin-tuntsov-02,rahvar-11,dorazio-distefano-20,dorazio-distefano-18,kelley-21,ingram-21}; we use the generalized definition instead.}. Self-lensing is often considered to be a minor effect because lensing distortions are proportional to the distance from the lens to the source. 
The optical depth for self-lensing, or probability to produce strong magnification and distortion, is of order $\sigma^2/c^2$, 
where $c$ is the speed of light and $\sigma$ is the velocity dispersion of the dynamical system containing the lens and the source, which is usually small compared to $c$. 

Our motivation to consider self-lensing of GWs is that the orbital velocity between the GW source and the lens may be high in some environments, such as in galaxy nuclei near the central massive BH, where BBH mergers may be accelerated by dynamical relaxation processes or the interaction with a quasar accretion disk of an AGN. In addition, future improvements in the signal-to-noise ratio of GW events may allow the detection of small lensing effects at impact parameters larger than the lens Einstein radius or lens masses lower than the merging BHs (for which diffraction reduces the gravitational lensing effects), hence increasing the lensing detection probability. Since next generation detectors such as Einstein Telescope (ET) and Cosmic Explorer (CE) are expected to detect a large amount of events, \mbox{$\sim (10^5-10^6)\, {\rm events}/{\rm year}$} \cite{maggiore-20-ET-case,evans-21-CE-white-paper}, self-lensing can be detected for lensing probabilities as small as $\sim10^{-6}$.
Upper limits on the number of GW self-lensing events may help constrain the GW emission environment. 

Lensed GWs are expected to be 
detected either as multiple images separated by a time delay 
or as a distortion of the waveform caused by wave optics effects
(diffraction and interference) \cite{nakamura-98, takahashi-03}.
Generally, wave optics effects are important 
when the wavelength is longer than the Schwarzschild radius corresponding to the lens mass, and close to caustics where they also become important even for high lens masses \cite{lo-25-high-magnification}. The time delay can be used as a more precise criterion: for a point mass lens at least, wave optics is important when the time delay between the lensed images is shorter than the GW period \cite{BU-22}.
Therefore, the  
geometric optics (GO) limit for GWs (when the waves can be approximated as rays) is valid  
when the time delay between the lensed images is longer than the period of the GWs.
Although lensed GWs have not yet been detected \cite{hannuksela-19,dai-20,LVK-O3a-lensing, LVK-O3ab-Lensing,janquart-23} (but see \cite{231123,2025arXiv250821262S}), the first lensed event caused by a  foreground lens (a galaxy or galaxy cluster) may be observed during 
the fourth LVK observing run (O4) \cite{ng-18,li-18,oguri-18,xu-ezquiaga-holz-22,mukherjee-21,wierda-21}. The possibility of self-lensing needs to be taken into account for short time delays corresponding to stellar, intermediate-mass BHs (IMBHs) or supermassive BHs (SMBHs) as lenses.   

The use of self-lensing to learn
about the environment of the source has been studied before with optical sources, for example in stellar and planetary binary systems,
\cite{maeder-73,mao-paczynski-91,gould-loeb-92,witt-94,gould-binary-95,marsh-01,beskin-tuntsov-02,rahvar-11,dorazio-distefano-20},
or in binary SMBHs at the heart of AGN disks
\cite{dorazio-distefano-18,kelley-21,hu-20-spikey,ingram-21}. 
Self-lensing studies of GWs have focused on NCs and AGN disks, where a BBH orbits around and is lensed by the central SMBH \cite{kocsis-13,gondan-kocsis-22,dorazio-loeb-20,yu-21,toubiana-21,gondan-kocsis-22,zhang-chen-23}. In the latter case, BBHs are thought to be dragged by dynamical friction into the AGN disks 
around the SMBH by the gas dynamical friction 
\cite[e.g.,][]{oleary-09,mckernan-12,bellovary-16,stone-17,bartos-17,mckernan-18,ostriker-83,syer-91,artymowicz-93,levin-07}. 

In this study, we explore different environments and dynamical situations where the velocities are high, or where the self-lensing probability may be enhanced above the simple estimate of optical depths based on the scaling $\sigma^2/c^2$. 
We focus on GWs in the presently accessible LVK frequency range $f\simeq(10-10^3)\,{\rm Hz}$ and  for the future ET $f\simeq(1-10^3)\,{\rm Hz}$.
In Sec.~\ref{sec:probability}, we obtain the probability of self-lensing in different astrophysical environments. 
We consider dynamical interactions in star clusters, both with and without a massive central BH, and in AGN disks. In Sec.~\ref{sec:imprint}, we analyze the signature of lensing 
in each environment. 
The results are discussed in Sec.~\ref{sec:discussion}.

\section{Probability of self-lensing}
\label{sec:probability}

\begin{figure}
\centering
\includegraphics[width=\columnwidth]{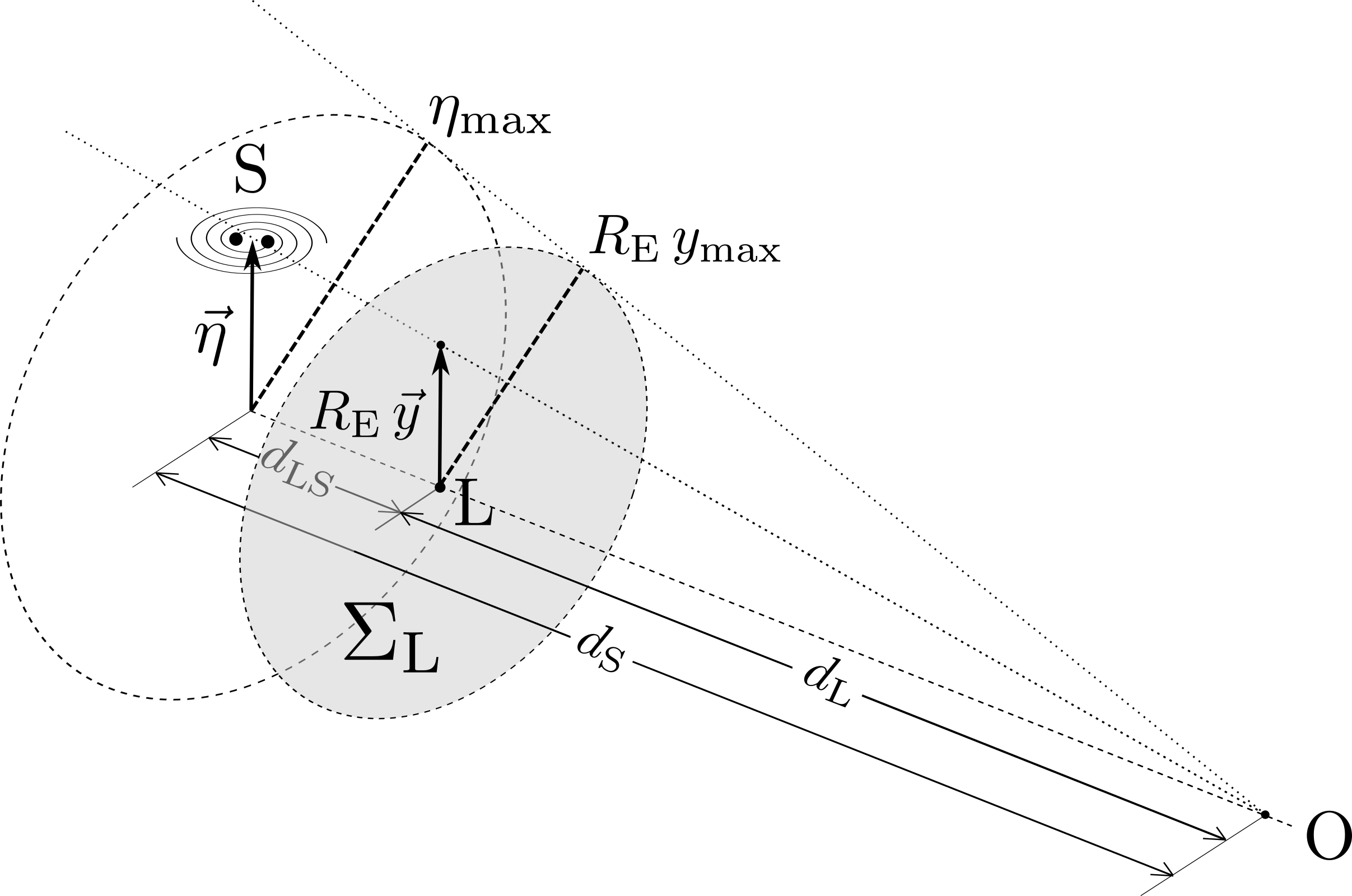}
\caption{Gravitational lensing diagram to illustrate our notation.
The source S is displaced from the observer-lens line-of-sight (observer at O, lens at L) by an impact parameter $\vec{\eta}=\re\, \vec{y}$ (Eq.~\eqref{eq:y_def}). The lens cross section $\Sigma_{\rm L}$ is shown as the grey circle of radius $\re \,\ymax$ at the lens plane. When the source is inside the empty dashed circle in the source plane, the GW signal has detectable lensing effects. 
}
\label{fig:geom-intro}
\end{figure}

We start by defining basic quantities used in this work. Gravitational lensing causes strong magnification and distortion when the impact parameter is close to the Einstein radius of a point mass lens of mass $\ML$, defined as 
\begin{align}
R_{\rm E} &= \left(\frac{4G\ML}{c^2}\frac{d_{\rm L}d_{\rm LS}}{d_{\rm S}}\right)^{1/2},
\label{eq:RE1}
\end{align}
where $G$ is the gravitational constant, $c$ is the speed of light in vacuum, $d_{\rm S}\,(d_{\rm L})$ is the distance from the observer to the source (lens), and $\dLS$ is the distance between the source and the lens. All distances are defined as angular diameter distances \cite{schneider-92}. 
We use the thin lens approximation, where $R_{\rm E}$ is small compared to $d_{\rm L}$ and $\dLS$, and lensing depends only on the projected surface density.
In the case of self lensing, $d_{\rm LS}\ll d_{\rm S}$ and $d_{\rm L}\simeq d_{\rm S}$, leading to
\begin{equation}
R_{\rm E} \simeq \left(\frac{4G\ML d_{\rm LS}}{c^2}\right)^{1/2} = \left(2R_{\rm L}\, d_{\rm LS}\right)^{1/2},
 \label{eq:RE}
\end{equation}
where $R_{\rm L} = 2G\ML/c^2$ is the Schwarzschild radius 
corresponding to the 
mass $\ML$.

The distances $d_{\rm S}$, $d_{\rm L}$, $\dLS$ are cosmological, therefore the observed masses of the lens and the source are redshifted and become respectively the redshifted masses $\ML (1+z)$ and $\MS (1+z)$. Here  $z$ represents the cosmological redshift and in our case both lens and source are at the same distance.

The dimensionless source position, sometimes called impact parameter, is defined as 
\begin{equation}
\vec{y}=
\frac{d_{\rm L}}{d_{\rm S}}
\frac{\vec{\eta}}{\re} \simeq
\frac{\vec\eta}{\re}
~,
\label{eq:y_def}
\end{equation}
where $\vec{\eta}$ is the displacement of the source from the lens-observer line-of-sight, as shown in Fig.~\ref{fig:geom-intro}. 

The cross section $\Sigma_{\rm L}$ behind which we observe the effect of a point mass lens is determined by the maximum impact parameter $\ymax$ for which lensing is detectable,
\begin{align}\label{eq:cs}
\Sigma_{\rm L}&=\pi \ymax^2 \re^2,\\
&=2\pi \ymax^2 R_{\rm L}\, \dLS ~.
\label{eq:cross-section}
\end{align}
As shown in Fig.~\ref{fig:geom-intro}, if the source is behind the projection of the lens cross section, the GW signal displays detectable lensing effects.

As detailed in Sec.~\ref{sec:detectability}, $\ymax$ is the maximum value that  $y$ can have for lensing to be detectable. 
Its value 
depends on the signal-to-noise ratio of the observed event, which we discuss in more detail in Sec.~\ref{sec:imprint}. 

The lensing optical depth, $\tau$, is the fraction of the lens plane area filled by the Einstein rings of all the point masses \cite{schneider-92, narayan-96}. It can be generalized by multiplying the Einstein radii by $\ymax$, which defines the threshold for lensing detection, to obtain an optical depth for a lensing detection.
The probability of self-lensing is $1-e^{-\tau}$, which for $\tau\ll1$ equals $\tau$, which we use in this work.

For a BBH merger source that orbits a (massive) BH that acts as a lens, the optical depth $\tau$ for a random orientation is \cite{gould-binary-95}
\begin{align}
\tau
&= \frac{\Sigma_{\rm L}}{4\pi \bar d_{\rm LS}^2}
=\frac{1}{2} \frac{\rs}{\bar d_{\rm LS}}\, \ymax^2 \label{eq:individual-prob1}\\
&= \frac{Gm_{\rm L}}{c^2 \bar d_{\rm LS}}\, \ymax^2,\label{eq:individual-prob2}\\ 
&\simeq \frac{v_{\rm orb}^2}{c^2 }\, \ymax^2,
\label{eq:individual-prob-scaled}
\end{align}
where $v_{\rm orb}$ is the time-averaged orbital velocity of the source around the lens at an average distance $\bar d_{\rm LS}$. This expression shows that self-lensing is important for sources moving with relativistic velocities in the potential of the lens. 

\begin{figure}
\centering
\includegraphics[width=\columnwidth]{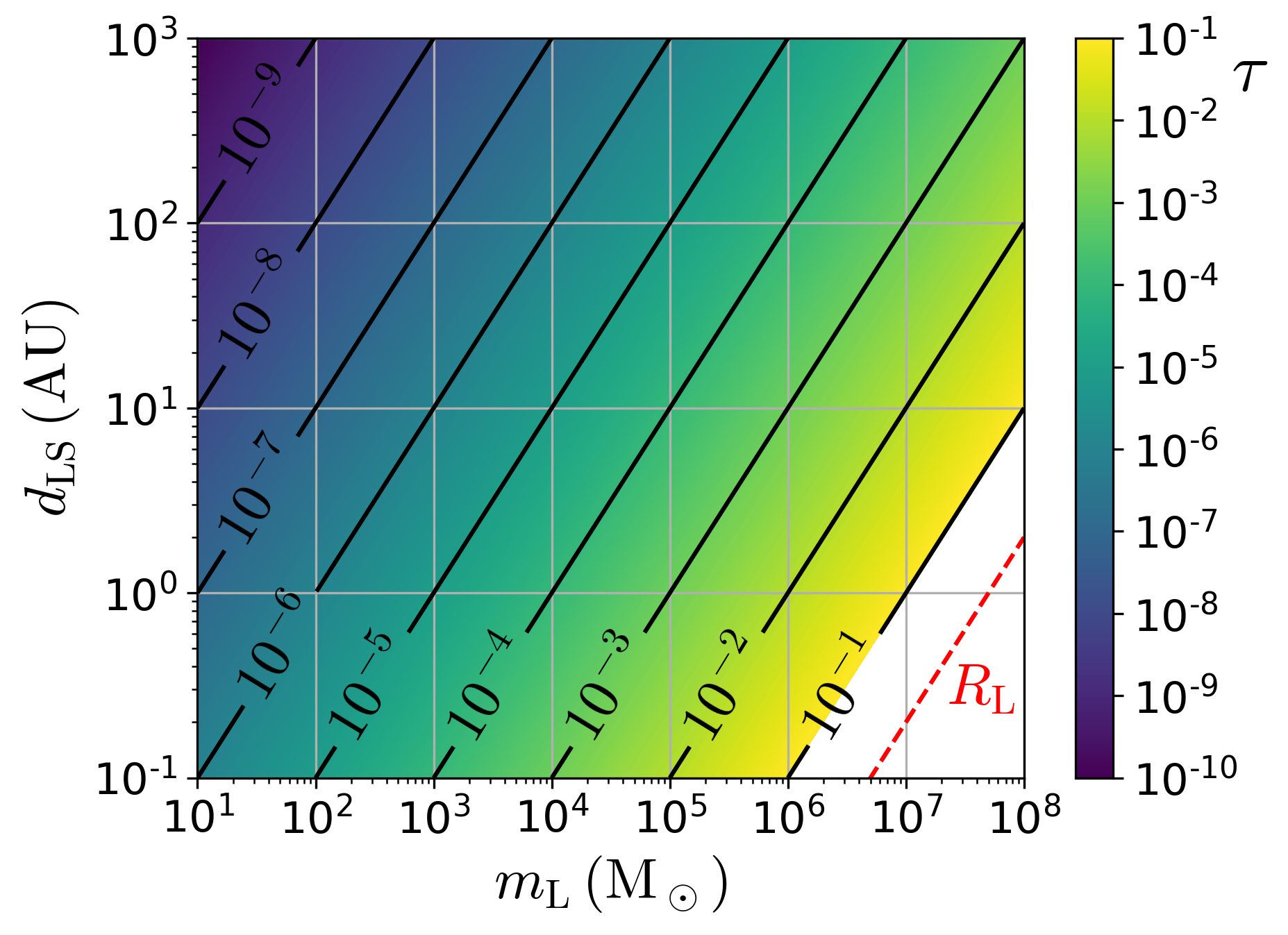}
\caption{
Self-lensing optical depth $\tau$ of a BBH merger lensed by a third object, as a function of the mass of the lens $\ML$ and the separation between the lens and the source $\dLS$, following Eq.~\eqref{eq:individual-prob-units}. 
We consider the LVK sensitivity, where $\ymax\sim 1$. The Schwarzschild radius of the lens, $\rs$, is shown as a dashed red line. 
}
\label{fig:pl-2d}
\end{figure}

It is useful to express $\tau$ in terms of characteristic units (solar masses $\msun$, and astronomical units $\au$),
\begin{equation}
\tau \simeq 10^{-7} 
\left( \frac{\ML}{10 \,\msun} \right)
\left( \frac{\au}{\dLS} \right)
\ymax^2,
\label{eq:individual-prob-units}
\end{equation}
as shown in Fig.~\ref{fig:pl-2d}.

In the next sections we quantify $\tau$ for self-lensing in six scenarios in three different environments: lensing by a stellar-mass BH in star clusters, lensing by the massive BH in nuclear clusters, and AGN disk.

\subsection{Lensing by a member of a star cluster}
\label{sec:prob-sigma}
We first consider the probability that a BBH merger is lensed by a star or BH in a star cluster. 
The signal of a compact BBH coalescence
is a short transient, compared to cluster orbital periods, so we can approximate its position as fixed. 

We start by considering the simple model of a singular isothermal sphere for the smoothed cluster mass distribution. The deflection angle is $4\pi \sigma^2/c^2$, so the cluster Einstein radius is 
$R_{\rm E,clus}=4\pi\, (\sigma/c)^2\, \dLS$ (for $\dLS\ll d_{\rm S}$). Here $\sigma$ is the one-dimensional velocity dispersion, which for the singular isothermal sphere relates to the circular velocity $V_{\rm c}$ as $\sigma=V_{\rm c}/\sqrt{2}$. In the singular isothermal sphere, the surface density is proportional to the inverse projected radius, $\Sigma\propto R^{-1}$, and therefore $\Sigma=\bar\Sigma/2$, where $\bar\Sigma$ is the mean surface density within $R$. Hence, $\Sigma=\Sigma_{\rm crit} R_{\rm E,clus}/(2R)$, where $\Sigma_{\rm crit}$ is the critical surface density for lensing \citep[e.g.,][]{TOG84,schneider-92}. If the smooth mass is all due to the stellar-mass microlenses, and because the mean $\Sigma/\Sigma_{\rm crit}$ inside the point mass Einstein radius is one, the optical depth is
\begin{equation}
    \tau= {\Sigma\over \Sigma_{\rm crit}}\ymax^2 = 2\pi\, {\sigma^2\over c^2}\, {\dLS\, \ymax^2\over R} ~.
\end{equation}
Note that the average optical depth for a source that is distributed in the same way as the smoothed mass in the singular isothermal sphere diverges logarithmically when integrating over the source distance to the lens, because the optical depth is proportional to $\dLS$ while the density falls off as the squared distance from the cluster center. An average lensing optical depth for all sources distributed in the same way as lenses can be computed in a stellar system of finite mass. As an example, Appendix \ref{appendix:general-probability} shows that the result for the mean optical depth $\bar\tau$ in a Plummer model is
\begin{equation}
    \bar\tau = 4\pi\, {\sigma^2 \ymax^2\over c^2}~. 
\end{equation}
As illustrated by the singular isothermal sphere case, the average optical depth has an important contribution from sources that are far from the cluster center compared to the Plummer model scale radius.

However, not all the objects that give rise to the potential are detectable lenses. The majority of the mass of the cluster is in the form of low mass stars, while for the GW sources we consider (BBH mergers), the detectable lenses are stellar-mass BHs, which constitute a few percent of the total mass for a canonical stellar initial mass function. As we will show in Sec.~\ref{sec:imprint}, it is necessary that $\ML\gtrsim m_{\rm S}$ (Eq.~\eqref{eq:ml-ms-ratio-LVK}) for detectability, where $m_{\rm S}$ the mass of the final remnant of the source. For high signal-to-noise ratio ($\snr$) events in ET, the threshold may be reduced to mass lenses lower than $\ML$ as shown in Sec.~\ref{sec:detectability}, although one should take into account that this result is optimistic and limited by degeneracies with other waveform distortions.

\begin{figure}
\centering
\includegraphics[width=\columnwidth]{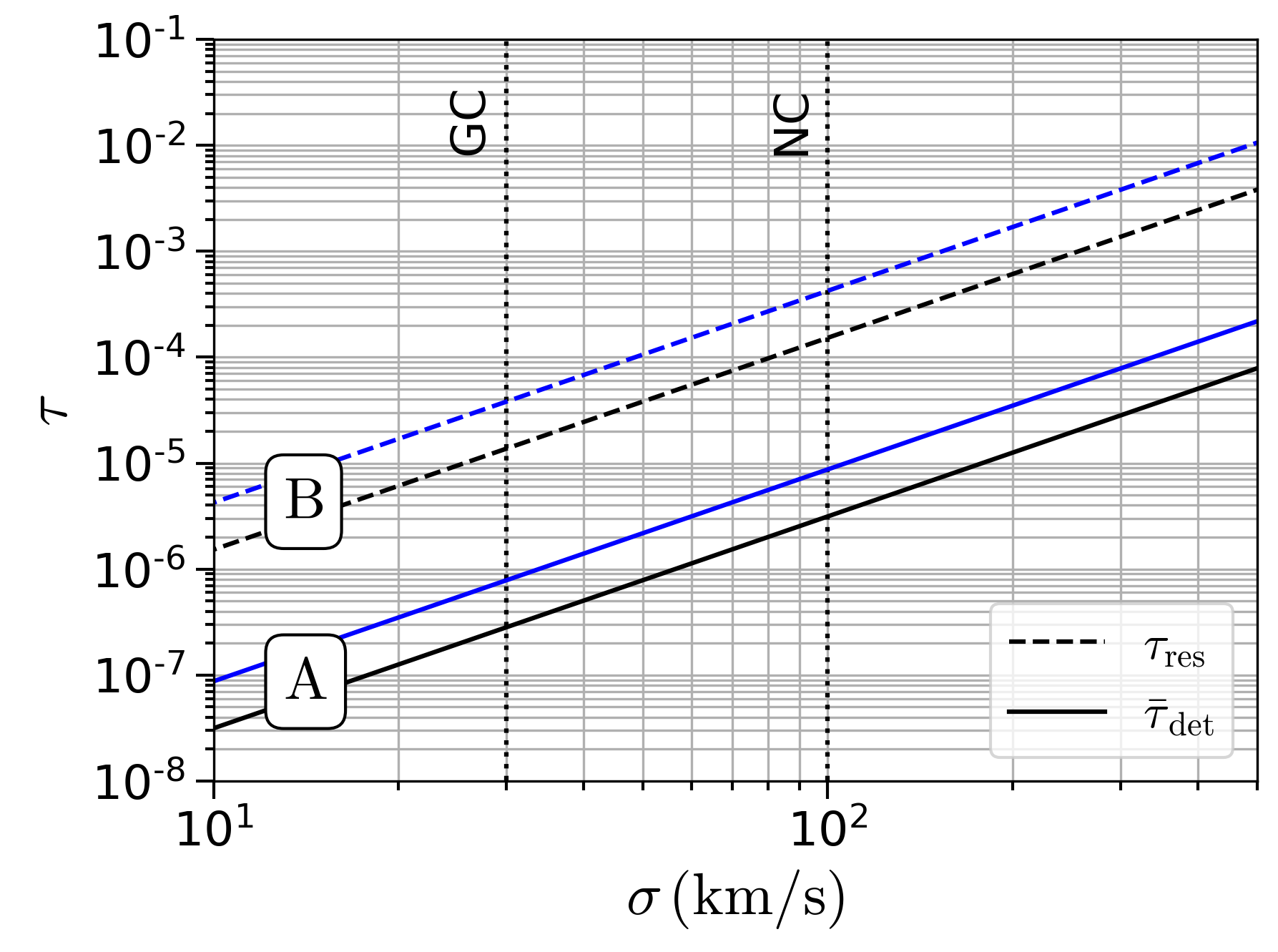}
\caption{Self-lensing probability $\tau$ as a function of $\sigma$, the star cluster velocity dispersion. The optical depth for self-lensing by a member of the star cluster (A), $\bar\tau_{\rm det}$ (Eq.~\eqref{eq:optical_depth_general_plummer_BH}), is shown as solid lines, assuming $f_{\rm L}=0.05$, $f_{r_0}=0.1$. Self-lensing of resonant interactions leading to a GW capture (B), with optical depth $\tau_{\rm res}$ (Eq.~\eqref{eq:prob-resonant}), is shown as dashed lines. Black (blue) lines correspond to LVK (ET), where we take respectively $\ymax\simeq1.5$ and $\ymax\simeq2.5$ (as detailed in Sec.~\ref{sec:detectability})
for $\MS=10\,\msun+10\,\msun$. The dotted vertical lines show a representative value of $\sigma$ for globular clusters (GC) and nuclear star clusters (NC).
}
\label{fig:pl_sigma0}
\end{figure}

The optical depth for the BBH merger sources in a cluster for detectable lensing
is (Eq.~\eqref{eq:optical_depth_general_plummer_BH_appendix}):
\begin{equation}
\bar\tau_{\rm det} \simeq10^{-7} \, 
\frac{f_{\rm L}}{0.05}\,\frac{0.075}{f_{r_0}} 
\left(\frac{\sigma}{30\,\kms}\right)^2 \ymax^2,
\label{eq:optical_depth_general_plummer_BH}
\end{equation}
where $f_{\rm L}=M_{\rm L}/M_{\rm clus}$ is the mass fraction of the detectable lenses (i.e., BHs) relative to the total cluster mass, 
and $f_{r_0} = r_{0, \rm L}/r_{0, {\rm clus}}$ is the (scale) radius of the detectable lens population relative to the scale radius of the cluster. The value of $\bar\tau_{\rm det}$ provides an estimate of the probability of self-lensing in different environments, as seen in Fig.~\ref{fig:pl_sigma0}. These can be GCs ($\sigma \sim 10-30 \, \kms$), NCs ($\sigma \sim 30-100\,\kms$) or galaxies as a whole ($\sigma \gtrsim 100 \, \kms$; $f_{r_0}=1$).
We note that in galaxies the BHs are not centrally concentrated ($f_{r_0}=1$), such that the highest probability for self-lensing is in NCs: $\bar\tau_{\rm det}\sim 10^{-6}-10^{-5}$ for $\sigma\simeq100\,\kms$ and $1.5\lesssim\ymax\lesssim2.5$ (the values of $\ymax$ are obtained in Sec.~\ref{sec:detectability}). 

\subsection{Resonant interactions in dense stellar systems}
\label{sec:resonant}
\begin{figure} 
\centering
\includegraphics[width=0.9\columnwidth]{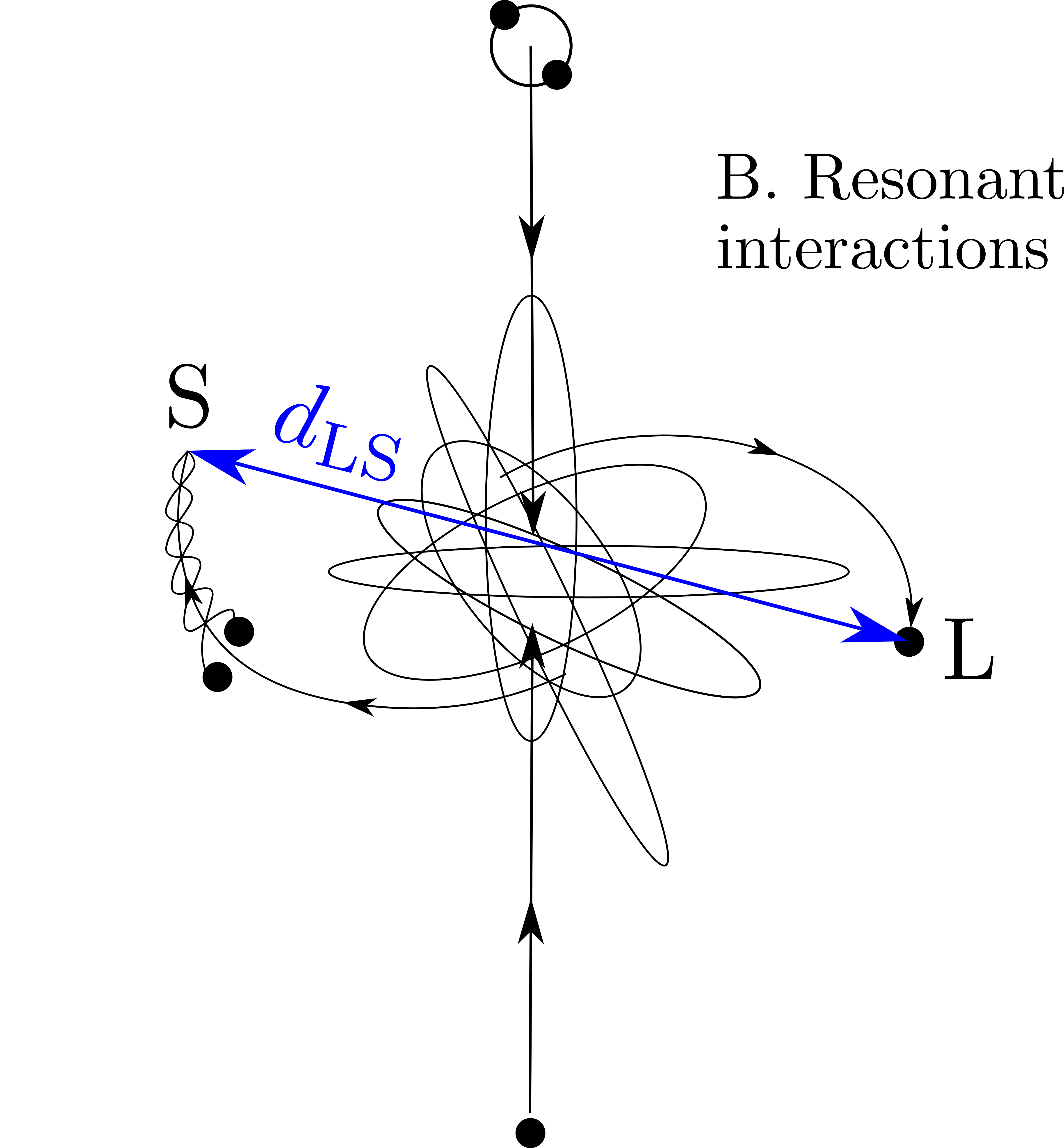}
\caption{Diagram of a BBH-single BH interaction leading to a GW capture. A BBH (above) and a single BH (below) encounter each other, interact chaotically (resonant encounters), interchanging its components. We show the outcome we are interested in: the formation of a new BBH that merges quickly (left, S) close to the remaining single BH (right, L). This single BH can act as a lens. Not to scale: in reality, $\dLS\sim a_0$, the initial semi-major axis of the binary.
}
\label{fig:dynamical-3body}
\end{figure}

Here we consider the probability of lensing of a GW capture (defined as an unbound orbit becoming bound by GW emission) occurring during strong gravitational interactions, or ``resonant interactions" \cite{1983ApJ...268..319H} between BBH-single BH \cite{samsing-20} or BBH-BBH \cite{Zevin2019,marin-pina-gieles-25}. We are interested in the GW capture and coalescence while the resonant interaction is ongoing, such that a third object can be close and act as a lens, as illustrated in Fig.~\ref{fig:dynamical-3body}. 

From energy conservation arguments\footnote{
Assuming all three distances are the same and assuming a parabolic encounter, it follows that the separation between each BH is then $3a_0$ and $d_{\rm LS}=1.5\sqrt{3}a_0\simeq2.6a_0$. The separation is therefore  $d_{\rm LS}\gtrsim2.6a_0$. } (see also \cite{samsing-14}) it follows that the typical distance between the lens and binary during a resonant interaction is $\dLS\simeq a_0$, where $a_0$ is the initial semi-major axis of the binary before the interaction. 
The intermediate-state binaries have an approximately  thermal distribution of the eccentricity $e$, uniform in $e^2$, and those that result in merger are aided by a very high value of the eccentricity when they become bound, $e\gtrsim 0.9999$. 
We can obtain the optical depth from $\dLS$ (Eq.~\eqref{eq:individual-prob-units}) for different astrophysical environments.
For $a_0\sim 0.1\,\au$ and $\ML=10\,\msun$,
we obtain that $\tau \simeq 2\times 10^{-6}$ for $\ymax \simeq 1.5$ in LVK, $\tau \simeq 6\times10^{-6}$ for $\ymax \simeq 2.5$ in ET. 

To compare this scenario to $\bar\tau_{\rm det}$ by a member of a star cluster, we express $a_0$ in terms of $\sigma$ of the cluster: the most probable $a_0$ that can lead to a GW capture during an interaction is the smallest semi-major axis the dynamically active binary can have before it is ejected in a recoil after the interaction \cite{antonini-rasio-16,antonini-gieles-20}: 
$a_{\rm ej}=0.2G \ML / (6 v_{\rm esc}^2)$, where $v_{\rm esc}$ is the escape velocity from the center of the cluster and we assumed that all BHs involved have mass $\ML$.
For the Plummer model, $v_{\rm esc}^2/\sigma^2 = 64/\pi $ such that 
Eq.~\eqref{eq:individual-prob2} with $d_{\rm LS} = a_{\rm ej}$, we find
\begin{equation}
\dLS\simeq 0.016
\,{\au} \,\left(\frac{\ML}{10\,\msun}\right)\,\left(\frac{30\,\kms}{\sigma}\right)^2, 
\label{eq:dls-sigma}
\end{equation}
and the optical depth can be written as
\begin{equation}
\tau_{\rm res} \simeq 6\times 10^{-6} 
\left( \frac{\sigma}{30\,\kms} \right)^2\ymax^2 . 
\label{eq:prob-resonant}
\end{equation}
GW captures during resonant interactions constitute approximately 10\% of all dynamically formed BBH mergers in star clusters \cite{2018PhRvD..98l3005R, antonini-gieles-20}. Combined with the fact that $\tau_{\rm res}$ is $\sim60$ times higher than $\bar\tau_{\rm det}$ in the same cluster, from Eq. (\ref{eq:optical_depth_general_plummer}), 
self-lensing of a GW capture in a resonant interaction is $\sim6$ times more likely than self-lensing of any other BBH merger by another BH in the cluster.

A comparison between $\bar\tau_{\rm det}$ (Eq.~\eqref{eq:optical_depth_general_plummer_BH}) and $\tau_{\rm res}$ (Eq.~\eqref{eq:prob-resonant}) is shown in Fig.~\ref{fig:pl_sigma0}. As we will see in Sec.~\ref{sec:imprint}, the imprint in both cases is a chromatic amplification in the diffraction regime.

\subsection{Single-single GW capture near a massive BH}
\label{sec:prob-GW-capture}

\begin{figure}[!t]
\centering
\includegraphics[width=0.6\columnwidth]{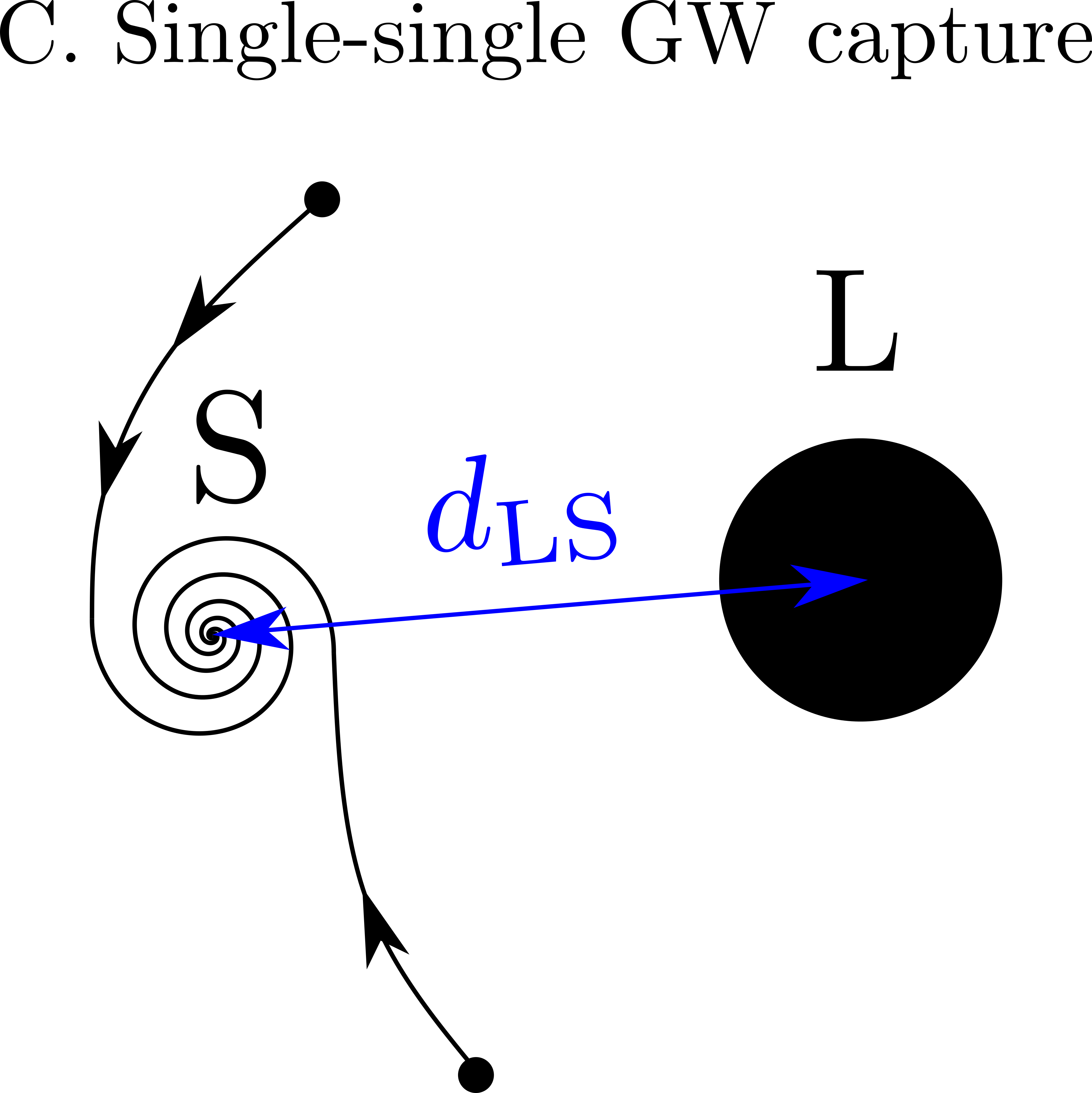}
\caption{Diagram of a hyperbolic (single-single) encounter leading to a GW capture between two stellar-mass BHs close to a massive BH. The interacting BHs can end up merging and being lensed by the massive BH. Not to scale.}
\label{fig:GW-capture-near-BH}
\end{figure}

NCs can have a population of stellar-mass BHs orbiting near a central massive BH. The stellar-mass BHs develop a Bahcall-Wolf cusp \cite{1976ApJ...209..214B} via relaxation, leading to densities high enough that GW captures can occur in (hyperbolic) single-single interactions \cite{oleary-09,2023MNRAS.518.5653B}, as illustrated in Fig.~\ref{fig:GW-capture-near-BH}.  

The probability of self-lensing in this case is determined by  the optical depth given the distribution of single-single GW captures lensed by the massive BH. We refer the reader to Appendix \ref{appendix:gw-capture} for the complete derivation. We obtain (Eq.~\eqref{eq:prob-ratio-approx}) that the probability
\begin{align}
\tau \simeq 4\times 10^{-5}\ &\left( \frac{\sigma}{30\,\kms}\right)^2
\left(\frac{10 \, m_{\rm L,c}}{M_{\rm clus}}\right)^{2/7} \times \nonumber \\
\times&
\left(\frac{m_{\rm L,c}}{10^6\,\msun}\right)^{4/7}
\left(\frac{m_{\rm S}}{20\,\msun}\right)^{-4/7}\, \ymax^2 
\label{eq:gw-capture}
\end{align}
depends on the mass of the cluster $M_{\rm clus}$, the mass of the BBH $m_{\rm S}$, and the mass of the BH central lens $m_{\rm L, c}$. 
This expression implies that $\tau$ is highest in NCs with the most massive central BHs, both in absolute terms and relative to $M_{\rm clus}$. 

To understand the dependence of $\tau$ on $m_{\rm L,c}$, one can fix $\MS=20\,\msun$, $M_{\rm clus}=10\,m_{\rm L, c}$, $\ymax=1$, and use the relation between $\sigma$ and $M_{\rm clus}$ (i) 
for NCs ($M_{\rm clus}\gtrsim3\times10^6\,\msun$): $\sigma\simeq30\,\kms\,(M_{\rm clus}/[10^7\msun])^{0.2}$; (ii) for GCs ($M_{\rm clus}\lesssim3\times10^6\,\msun$): $\sigma\simeq30\,\kms\,(M_{\rm clus}/[10^7\msun])^{0.5}$ \cite{hasegan-2005}. This leads to 
\begin{align}
&\tau\simeq 4\times 10^{-5} \left(\frac{m_{\rm L, c}}{10^6\,\msun}\right)^{0.97} \ymax^2,\,\, {\rm for}\, M_{\rm clus}\gtrsim3\times10^6\,\msun,
\label{eq:sscapt_high}
\\
&\tau\simeq 4\times 10^{-5} \left(\frac{m_{\rm L, c}}{10^6\,\msun}\right)^{1.57} \ymax^2,\,\, {\rm for}\, M_{\rm clus}\lesssim3\times10^6\,\msun,
\label{eq:sscapt_low}
\end{align}
which we use later in Fig.~\ref{fig:summary-probabilities}.

For the NC in the center of a Milky Way-like galaxy, $m_{\rm S} \simeq 20 \,\msun$, $\sigma\simeq 50\, \kms$, $m_{\rm L,c}\simeq  4\times10^6 \,\msun$, $M_{\rm clus}\simeq 2\times10^7\,\msun$ and $\ymax=1$, we obtain $\tau \simeq 3\times10^{-4}$. 

Another case of interest is an IMBH at the center of a GC. 
For $m_{\rm S} = 20 \,\msun$, $M_{\rm clus}= 10^5 \,\msun$, $m_{\rm L,c}= 10^4 \,\msun$, and $\sigma=10\,\kms$ the probability is $\tau\simeq 3\times 10^{-7}$. In these particular conditions, the low $m_{\rm L,c}$ and $\sigma$ make this an unlikely self-lensing case. 

The probability of single-single GW captures in NCs can reach high enough values $\tau\sim10^{-4}$ to be considered for next-generation detectors, where the number of detections will be large, \mbox{$\sim (10^5-10^6)\, {\rm events}/{\rm year}$}. Additionally, GW captures start with extreme eccentricity ($e\gtrsim0.999$ \cite{hansen-72}), such that there is residual eccentricity once the BBH enters the observable GW frequency range, a signature that can be combined with self-lensing to determine their origin.

\subsection{von Zeipel-Lidov-Kozai oscillations in star clusters with a central massive BH}
\label{sec:prob-LK}

\begin{figure}[!t]
    \centering
    \includegraphics[width=0.8\linewidth]{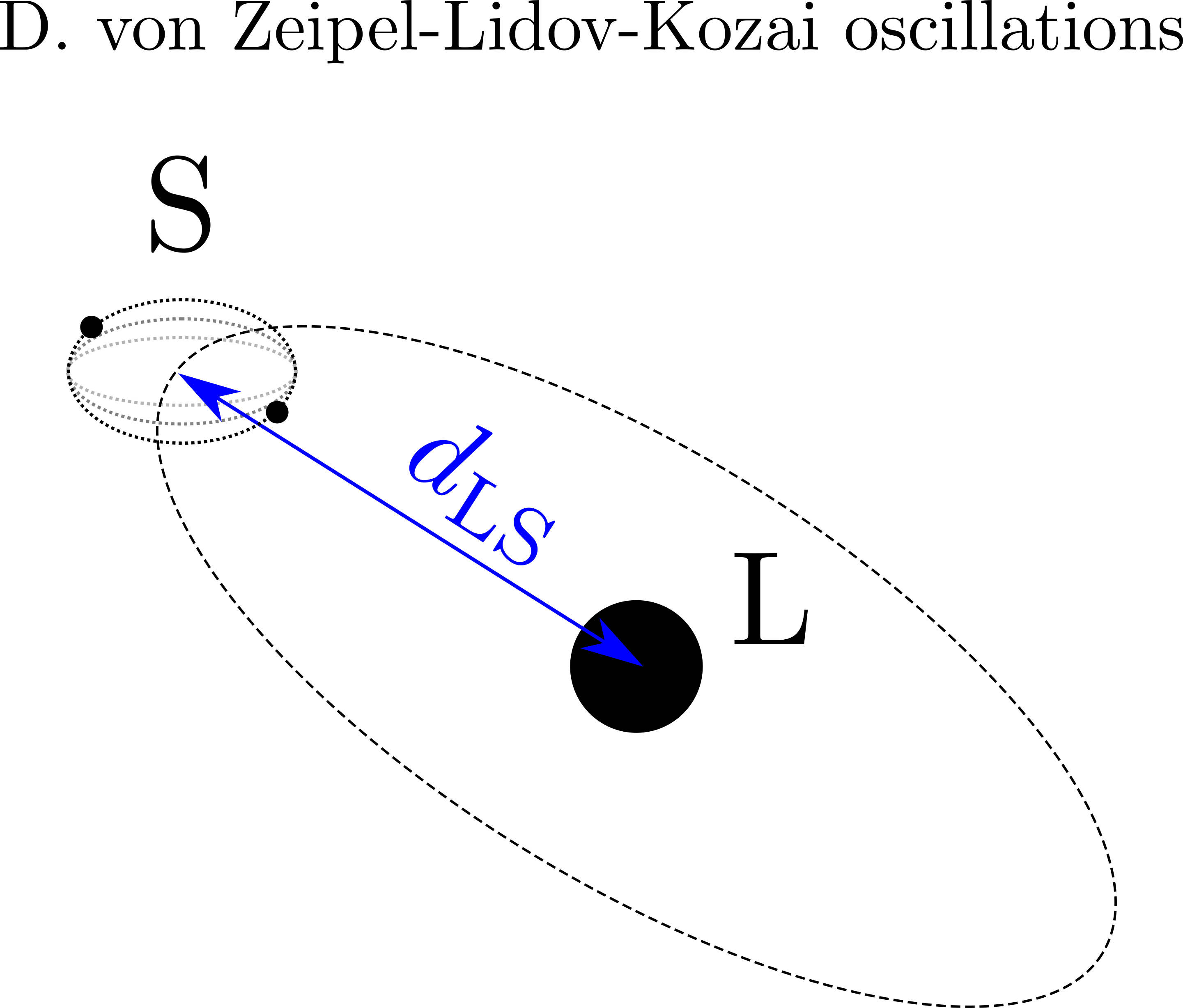}
    \caption{Diagram of a BBH orbiting a massive BH. The BBH can be influenced by the massive BH, leading to ZLK oscillations that alter the eccentricity of the BBH and trigger its merger. The ZLK effect becomes more important when the BBH orbital plane is significantly inclined with respect to the of the orbit around the massive BH. Not to scale.}
    \label{fig:ZLK}
\end{figure}

A central massive BH in a star cluster can perturb the eccentricity of the orbit of a stellar-mass BBH, leading to von Zeipel-Lidov-Kozai (ZLK) oscillations \cite{von-zeipel-10,lidov-62,kozai-62}. This perturbation can either disrupt the BBH or trigger its merger if the binary orbit eccentricity grows to near unity. 
If the BBH merges behind the massive BH,
lensing can occur as shown in Fig.~\ref{fig:ZLK}.

In Appendix~\ref{appendix:ZLK} we estimate how the optical depth depends on the mass of the central BH, guided by the Monte Carlo models of \cite{hoang-18}. There we show that $\tau$ mainly depends on $m_{\rm L,c}$:

\begin{equation}
\tau\simeq 3\times10^{-5} \left(\frac{m_{\rm L,c}}{10^6 \,\msun}\right)^{2/3} \ymax^2~.
\label{eq:ZLK-realistic}
\end{equation}
This optical depth is higher than the one of single-single GW captures for $m_{\rm L,c}\lesssim3\times 10^5\,\msun$ (Eq.~\ref{eq:sscapt_low}), while it is comparable to it for $m_{\rm L,c}\gtrsim3\times 10^5\,\msun$ (Eq.~\ref{eq:sscapt_high}). In the first case, self-lensing is most likely to come from ZLK mergers. In the latter case, 
the presence of eccentricity can identify the environment, because nearly all GW captures are expected to lead to eccentric waveforms, while for ZLK this is less common \cite{antonini-perets-12}.

\subsection{Central massive BH in an AGN disk}
\label{sec:AGN-central}

\begin{figure}
\centering
\includegraphics[width=\columnwidth]{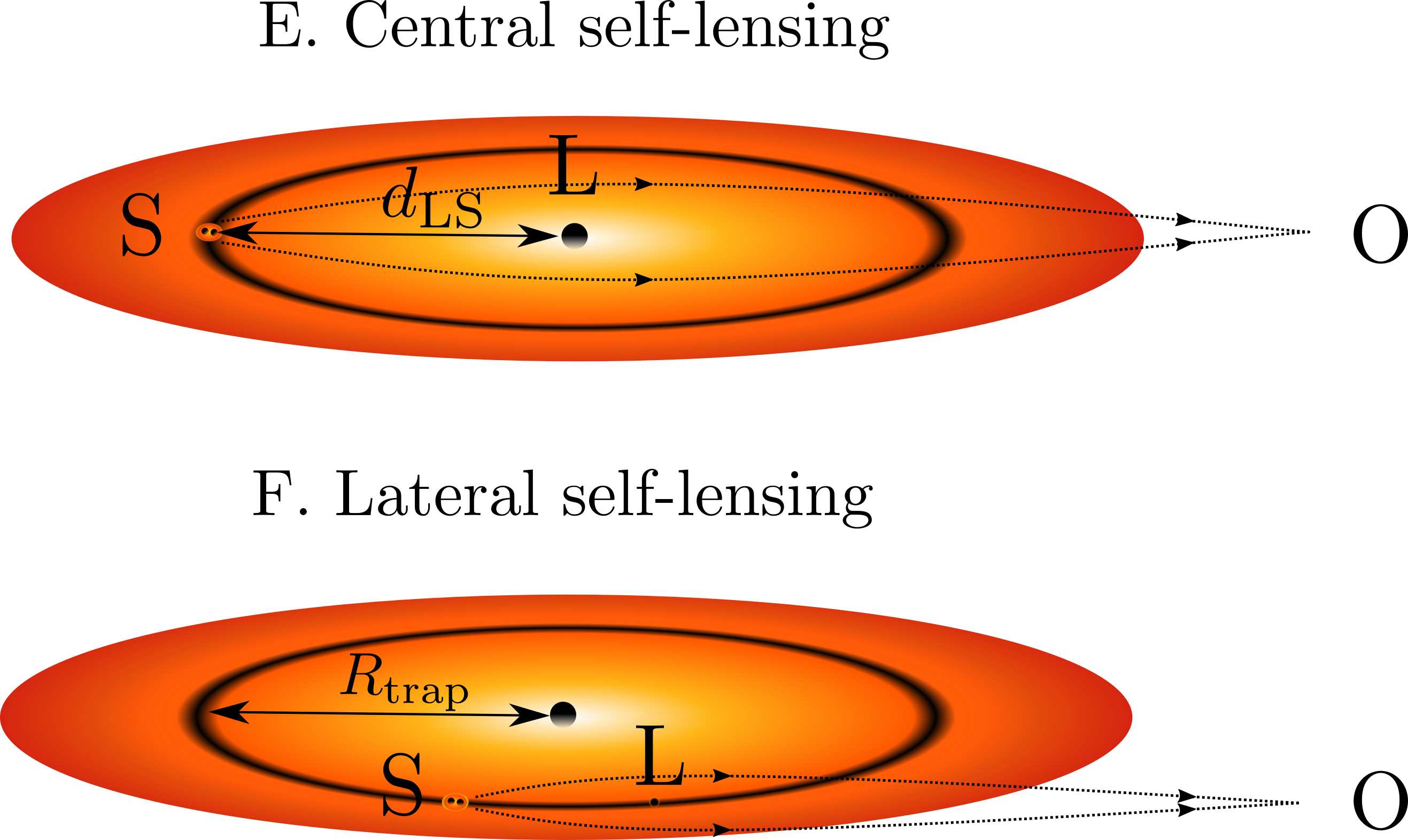}
\caption{Self-lensing of GWs from stellar-mass BBHs in the AGN disk. The BBH and other objects may be in a migration trap (dark orbit in the figure). \textbf{Top:} The BBH merger is self-lensed by the central SMBH. In this case the BBH merges behind the SMBH. \textbf{Bottom:} The BBH merger is self-lensed by another object in the disk (for example at the same orbit, in the migration trap). In this case, the BBH is lateral to, not necessarily behind, the SMBH. The source is at S, the lens at L and the observer at O, at a very large distance. Not to scale.}
\label{fig:AGN_graphic}
\end{figure}

AGNs in the centers of active galaxies host a central SMBH (or SMBH binary\footnote{In this work we consider a single SMBH for simplicity. If there were a SMBH binary instead, the lensing effect would look different (Sec.~\ref{sec:binary}). 
This is an interesting problem which is left for future work.
}) surrounded by an accretion disk. 
Stellar-mass BHs can be dragged into the gas disk by dynamical friction \cite{ostriker-83,syer-91,artymowicz-93,levin-07}. Once in the disk, they can migrate towards migration traps \cite{mckernan-12,bellovary-16,secunda-19}. It has been suggested that stellar-mass BBH mergers occur inside the migration traps in AGN disks (e.g.,~\cite{oleary-09,mckernan-12,bellovary-16,stone-17,bartos-17,mckernan-18}). Their GW signal can be lensed by the central SMBH, 
as shown in Fig.~\ref{fig:AGN_graphic} (top), which we consider in this subsection. They might also be lensed by another stellar-mass BH or IMBH inside the disk, shown in Fig.~\ref{fig:AGN_graphic} (bottom), which will be addressed in Sec.~\ref{sec:AGN-lateral}.

Compared to the case of lensing by the SMBH in the NC (subsections \ref{sec:prob-GW-capture}, \ref{sec:prob-LK}), there will be two main differences due to the presence of an accretion disk: 

(i) Assuming BBHs are formed in migration traps at tens of Schwarzschild radii from the central SMBH, the distance from the BBH merger source to the SMBH lens is smaller, which increases the self-lensing probability; 

(ii) 
Self-lensing in accretion disks can only be observed 
from nearly edge-on accretion disks due to geometry, since the BBH inside the disk needs to be behind the SMBH lens during the merger. 
This restriction, nevertheless, can provide unique characteristics to the system.  
Because the BHs orbit in a thin ring around the SMBH, the BBH orbital plane is preferentially aligned with the one of the accretion disk \cite{king-05}.  The detected signal will therefore have a unique signature, linear $h_+$ polarization, that will enable us to distinguish this environment from others (Sec.~\ref{subsec:imprint-AGN}).

The BBH orbits around the massive BH, as shown in Fig.~\ref{fig:AGN_graphic} (top). By taking into account all the possible inclinations, the optical depth is given by Eq.~\eqref{eq:individual-prob1}, 
\begin{equation}
\tau \simeq \frac{1}{2}\left(\frac{\rs}{R_{\rm trap}}\right) \ymax^2~.
\end{equation}
The distance to the central object $\dLS=R_{\rm trap}$ can be significantly reduced if the BBHs accumulate in migration traps \cite{bellovary-16,secunda-19}. Those are located at radii $R_{\rm trap}$ which scale with the Schwarzschild radius of the central massive BH $R_{\rm L,c}$. 

If we take the closest migration trap reported in \cite{bellovary-16,secunda-19}, \mbox{$R_{\rm trap}\simeq 25 R_{\rm L,c}$}, the probability of self-lensing is \mbox{$\tau \simeq 0.02$}, i.e., $2\%$, independent on the lens mass. 
This probability is consistent with the results in \cite{gondan-kocsis-22}.
Other values found in literature are discussed in Sec.~\ref{sec:assumptions}.

\subsection{Lateral BH lens in an AGN Disk}
\label{sec:AGN-lateral}
Because the escape velocity in the migration trap of an AGN is extremely high, $v_{\rm esc} = c \,(R_{\rm L,c}/R_{\rm trap})^{1/2}\simeq(0.1-0.3)c$, for $R_{\rm trap}/R_{\rm L,c}\simeq10-100$, the BBH merger remnant remains close to the migration trap after it receives a relativistic  kick of a few $10^3\,\kms$. This allows the merger remnant to be involved in subsequent mergers, and it has been suggested that this hierarchical growth is a possible pathway to growing an IMBH in AGN disks \cite{mckernan-12,mckernan-14}. 
This leads us to consider an additional lensing configuration inside of AGNs: both source and lens are inside the disk, orbiting the SMBH at a certain distance. We note, however, that if the lensing BH is too massive it may clear the migration trap as in protoplanetary disks \cite{goldreich-tremaine-80} and prevent this scenario to happen.

The geometry of this lateral self-lensing case is represented in Fig.~\ref{fig:AGN_graphic} (bottom). Although the lens could be of stellar mass, most events are not likely to have a detectable lensing imprint (as explained in Sec.~\ref{sec:imprint}). The lateral lens may be an IMBH instead, formed through subsequent (hierarchical) BBH mergers, 
for which the lensing imprint will be a characteristic detectable interference pattern (see Sec.~\ref{sec:imprint}). We consider that the mass ratio between the lateral lens of mass $m_{\rm L, lat}$ with respect to the central BH of mass $m_{\rm c}$ is $m_{\rm L, lat}/m_{\rm c}\lesssim 10^{-2}$.

We assume both the BH lens and the BBH merger source are on the same orbital plane, which is randomly inclined with respect to the observer. They are in a circular orbit with radius $R=R_{\rm trap}\gg R_{\rm E, \, lat}$ around the SMBH. We take a sphere of radius $R_{\rm trap}$ and the angle $\theta$ defined as a polar angle with origin at the point of the sphere closest to the observer. Lensing will occur for half of the sphere, $\theta<\pi/2$. The probability distribution of $\theta$ is $\sin \theta \,d\theta$. The fraction of the sphere solid angle covered by the lens is $\pi \re^2\ymax^2/(4\pi R_{\rm trap}^2 \cos\theta)$.
For a single lens, the probability to find the source within this fraction of the sphere is
\begin{align}
\tau &= {1\over 2}\int_0^{\pi/2} d\theta\, \sin\theta
{\re^2\,\ymax^2 \over 4 R_{\rm trap}^2\cos\theta} \\
&=
{R_{\rm L,lat}\,\ymax^2 \over 2 R_{\rm trap}}\int_0^{\pi/2} d\theta\, \sin\theta ~,
\end{align}
whose solution is
\begin{align}
\tau &\simeq \frac{1}{2}\left(\frac{R_{\rm L, lat}}{R_{\rm trap}}\right) \ymax^2 \\
&\simeq 2 \times 10^{-5} \,\left(\frac{m_{\rm L, lat}}{10^3\, \msun}\right) \left(\frac{10^6\, \msun}{m_{\rm c}}\right)\ymax^2,
\end{align}
where in the latter we take $R_{\rm trap}=25 R_{\rm L,c}$, and $R_{\rm L, lat}$ 
corresponds to the Schwarzschild radius of the lateral lens. 

\begin{figure}
\centering
\includegraphics[width=\linewidth]{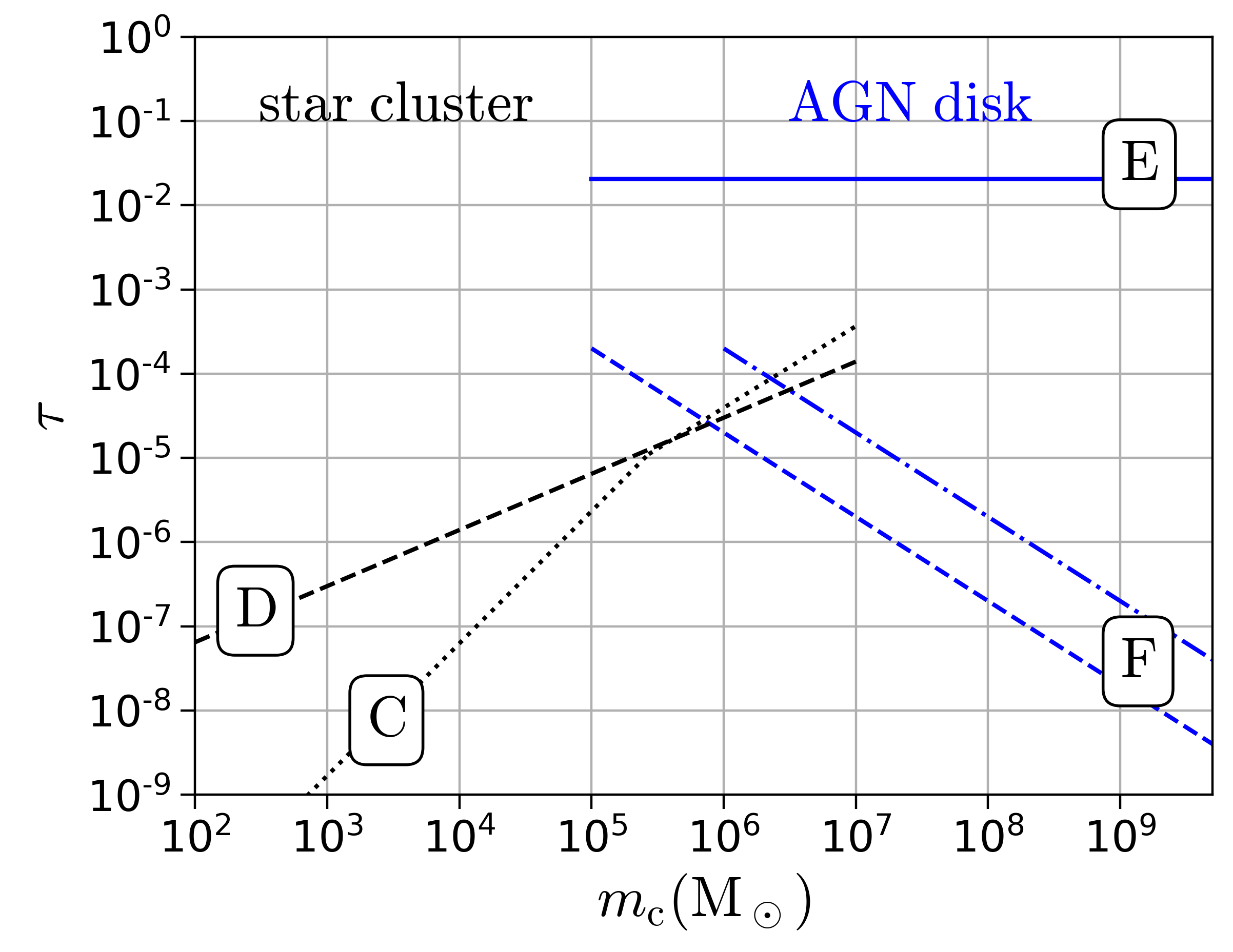}
\caption{Probability of self-lensing in environments with a massive central BH, as a function of its mass. 
The cases considered in star clusters (black lines) are the self-lensing of single-single GW captures near the massive BH (C) and self-lensing of ZLK mergers (D). For the single-single GW captures, the two slopes correspond to Eq.~\eqref{eq:sscapt_high} for $m_{\rm L,c}\gtrsim3\times 10^5\,\msun$ and Eq.~\eqref{eq:sscapt_low} for $m_{\rm L,c}\lesssim3\times 10^5\,\msun$. For the ZLK, we take the limiting value in Eq.~\eqref{eq:ZLK-realistic}. In both these cases the lens is the central BH ($m_{\rm c}$). AGN disks (blue lines) can have self-lensing by the central BH (E) or lateral lensing by an IMBH embedded in the disk (F), for which we plot two example values of the IMBH mass: $m_{\rm L,lat}=10^3\,\msun$ as dashed lines and $m_{\rm L,lat}=10^4\,\msun$ as dot-dashed lines, taking $R_{\rm trap}\simeq 25\,\rs$.} 
\label{fig:summary-probabilities}
\end{figure}

In Fig.~\ref{fig:summary-probabilities} we summarize the probability of self-lensing in the environments considered in Sec.~\ref{sec:prob-GW-capture}-\ref{sec:AGN-lateral}. As explained in the next section, they correspond mostly to interference and strong lensing imprints.

In summary, the AGN disk is the environment where the probability of self-lensing is highest. In star clusters, the presence of a massive BH also enhances the self-lensing probability for sources from (single-single) GW captures and ZLK mergers up to $\tau\sim 10^{-4}$.

\section{Imprint and detectability of self-lensing}
\label{sec:imprint}

\begin{figure*}[t]
\centering
\includegraphics[width=2\columnwidth]{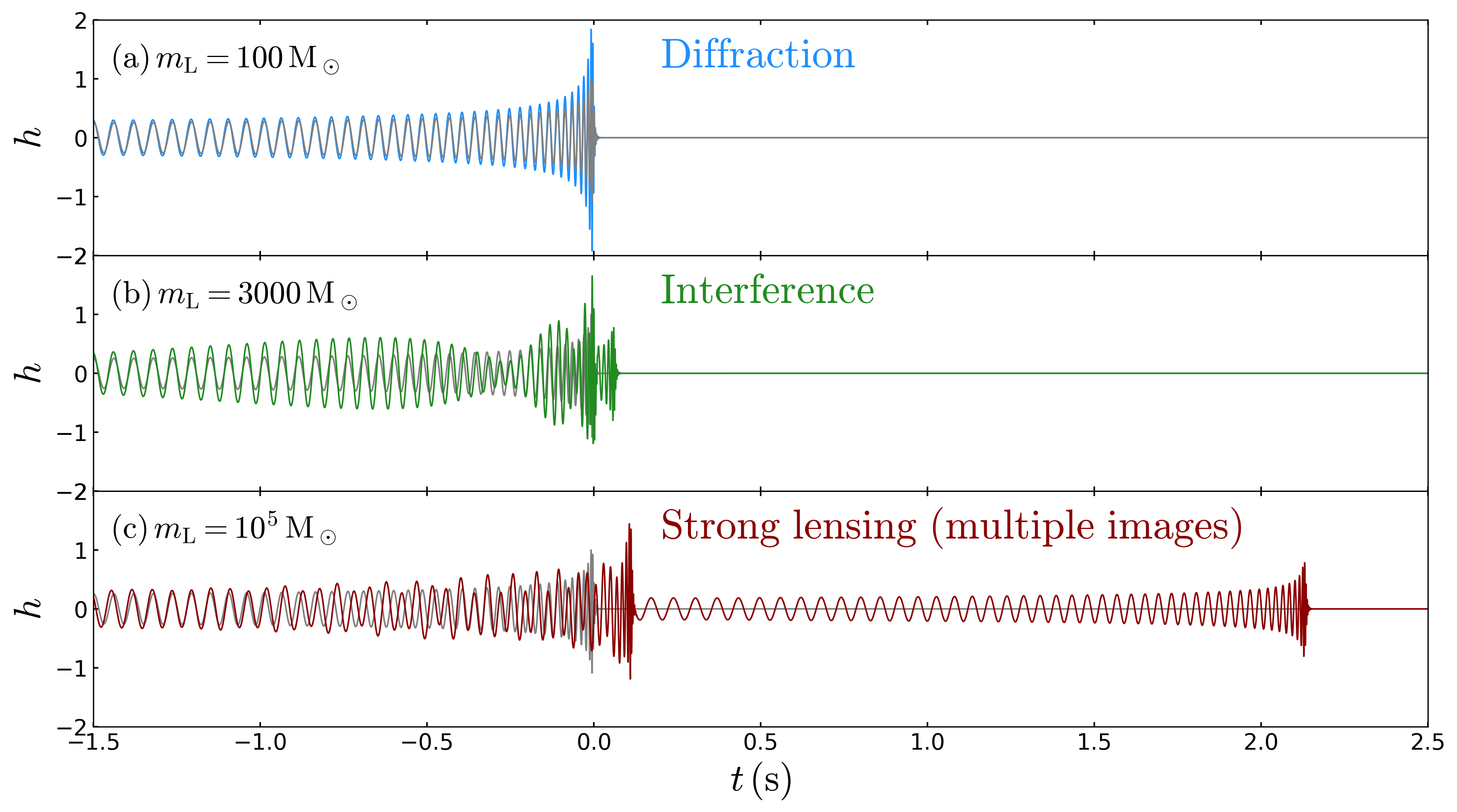}
\caption{Imprint of lensing on waveforms, presented as strain $h$ (arbitrary normalization) as a function of time $t$, for $y=0.5$ and 
lens masses ${\rm (a)}\,100\,\msun$, ${\rm (b)}\,3000\,\msun$, ${\rm (c)}\,10^5\,\msun$. The unlensed waveform, modelled for a $\MS=(10+10)\,\msun$ BBH merger source, is shown in grey for comparison with the lensed waveforms in colors, respectively in blue, green and red.
In the diffraction regime (a), the chromatic amplification is largest at highest frequencies (Eq.~\eqref{eq:transmission-factor-lowfreq}). When interference occurs (b), the emerging images overlap, producing a characteristic interference ``beating" pattern. Note that two images start to be seen at the end of the signal. For strong lensing (c), the images are well separated by a time delay (Eq.~\eqref{eq:time-delay}) with different amplifications (Eq.~\eqref{eq:magnif}, and residual interference only. 
}
\label{fig:lens_configuration_imprints}
\end{figure*}

In this section we discuss  the self-lensing effect (imprint) on a GW signal and its detectability.
The imprint is a distortion on the wave that in general depends on the lens mass, mass profile and the geometrical configuration between source, lens and observer. We use the point mass lens for this work. The binary components of the BBH source are non-spinning and non-precessing for simplicity.

We first review the types of imprints (diffraction, interference, strong lensing) by a point mass lens as a function of its parameters (\ref{sec:imprint-basics}). In \ref{sec:detectability}, we study the detectability of the imprint for the diffraction and interference range, while in \ref{sec:detectability-strong} we study the detectability of strong lensing. Finally, in \ref{subsec:imprint-environments} we discuss how the self-lensing imprint of the previously considered environments may be distinguished.

\subsection{Imprint by a point mass lens}
\label{sec:imprint-basics}

The detected lensed GW strain,
$\tilde{h} (f)$, 
is the product of the transmission factor $F$ and the unlensed strain $\tilde{h}_{\rm UL} (f)$ in the frequency domain:
\begin{equation}
\tilde{h} (f) = F(f,\ML,y)\, \tilde{h}_{\rm UL} (f)~.
\label{eq:strain_freq}
\end{equation}

The imprint of the gravitational lens on the GWs 
is given by the transmission factor $F$ \cite{deguchi-86a}, which for a point mass lens is 
\begin{equation}
F = e^{\frac{1}{2}\pi^2 \nu} e^{i \pi \nu  \ln (\pi\nu)} \,
\Gamma (1- i \pi \nu ) \,_1F_1 ( i \pi \nu; \,1; \,i \pi \nu y^2 ) ~, 
\label{eq:transmission-factor-full}
\end{equation}
with $\nu = 2 f \rs/c$, where $_1F_1$ is Kummer's confluent hypergeometric function and $\Gamma$ is the Gamma function.

For compact binary mergers, the frequency depends on the source mass and increases monotonically with time, reaching $f \simeq 1.2\times 10^4 (\msun/\MS) \, {\rm Hz}$ at merger \cite{maggiore-07}, where $\MS$ here is the final mass of the source remnant.

The lensing effects can qualitatively be separated 
into 
(a) diffraction, (b) interference and (c) strong lensing. The characteristic waveforms in the three regimes are shown in Fig.~\ref{fig:lens_configuration_imprints}.
Diffraction and interference are \textit{wave effects}, where the full wave optics formalism is required (\textit{wave optics lensing}). 
Strong lensing with multiple images can be treated in the geometric optics (GO) limit instead, which approximates the waves as rays. As we shall see in the following, some cases of interference can also be treated with the GO approximation.

\textbf{(a) Diffraction} imprints a frequency-dependent (\textit{chromatic}) amplification. 
It generally occurs when the wavelength $\lambda$ of GWs is comparable to, or larger than, the gravitational length of the lens ($2\rs$), although
close to caustics\footnote{Caustics are positions in the source plane where the magnification diverges for the ray (GO) approximation. In reality, the GO approximation breaks down close to them: wave optics becomes important, and it limits the maximum amplification. In the point mass lens the caustic is a point at $y=0$.} diffraction can be important for $\lambda\gtrsim8\rs\,y$, or equivalently $f \ML y\lesssim 10^4 \,{\rm Hz}\,{\msun}$ \cite[e.g.,][]{BU-22}. 
In the limit of low $f \ML y$, the transmission factor (Eq.~\eqref{eq:transmission-factor-full}) can be approximated as the asymptote
\begin{equation}
|F(\nu)| = \displaystyle{
\left( \frac{2 \pi^2 \nu}{1-e^{-2 \pi^2 \nu}} \right)^{1/2}}~.
\label{eq:transmission-factor-lowfreq}
\end{equation}
The imprint is a monotonic chromatic amplification of the signal. 
The higher the frequency, the more amplified the signal is. The approximation \eqref{eq:transmission-factor-lowfreq} is strictly valid up to $f \ML y\simeq 6250$ ($\nu_+$ in \cite{BU-22}).

\textbf{(b) Interference} imprints a characteristic pattern on the signal, which represents the interference between two emerging images in the GO approximation. The GO approximation starts from the point 
\mbox{$f \ML \yGO\gtrsim \, 1.25 \times 10^4 \, {\rm Hz}\, \msun$}. This threshold impact parameter, 
$\yGO$ \cite{BU-22}, can be related to the masses of the source and the lens:
\begin{equation}
\yGO = \frac{c}{8f \rs} 
\simeq \frac{\MS}{\ML},
\label{eq:yGO}
\end{equation}
where we used the frequency at the last stage of the coalescence (ringdown) 
$f\simeq 1.2 \times 10^4 \,{\rm Hz} \, (\msun/\MS)$ \cite{maggiore-07} to relate $\yGO$ to $\MS$. 
Additionally to the interference pattern, the signal is \textit{achromatically} (frequency-independent) amplified. 
As $f \ML y$ increases, the interference can be accompanied by the resolvability of the emerging images. In GW science, the images are practicably impossible to resolve in sky location. Instead, they are accurately resolved in time domain. The time delay between the two images is 
\begin{equation}
\Delta t \simeq 40\,{\rm s}\left(\frac{\ML}{10^6\,\msun}\right) \,y  ~,
\label{eq:time-delay}
\end{equation}
simplified with an approximation valid for $y\lesssim 0.5$ \cite{BU-22}.

\textbf{(c) Strong lensing} imprints multiple images of the same source that have an 
achromatic amplification each. These images (two in the point mass lens model), are separated both in sky location and in time: the latter can be seen in Fig.~\ref{fig:lens_configuration_imprints} (c). Once the time delay (Eq.~\eqref{eq:time-delay}) is longer than the time resolution of the GW detectors, the images are completely separate in time domain, and the imprint are two images. In the GO approximation, the transmission factor is \mbox{$F=(\sqrt{\mu_+} +\sqrt{\mu_-} e^{i2\pi f \Delta t-i\pi/2})\,e^{i\varphi_1}$}. Each image has an amplitude $\tilde{h}_\pm = \sqrt{\mu_\pm} \, \tilde{h}_{\rm UL}$ ($+$ for the first, brightest image and $-$ for the second, faintest one), where the magnifications $\mu_\pm$ are achromatic and only depend on $y$:
\begin{equation}
\mu_{\pm}=\frac{1}{4} 
\left( \frac{y}{\sqrt{y^2+4}} + \frac{\sqrt{y^2+4}}{y} \pm 2 \right).
\label{eq:magnif}
\end{equation}

In the transmission factor, the phase term $e^{i\varphi_1}$ corresponds to the first image, which has a topological (Morse) phase shift of $0$. The second image has a phase shift that depends on $f$ and $\Delta t$, as well as on a topological phase shift of $\pi/2$.

\subsection{Detectability limit for wave optics effects}
\label{sec:detectability}

Once we know the type of imprint we can expect, we need to know if it will be detectable. For wave optics effects (diffraction and interference), the lensed event is a single distorted signal that we can compare with unlensed templates.

The detectability of lensed signals depends on how much they deviate from existing unlensed template waveforms.
This deviation can be quantified using the Fitting Factor \cite{apostolatos-95}, which measures the agreement between the signal and the best-matching template. The Fitting Factor ranges from 0 (no match) to 1 (perfect match). 
For simplicity, we approximate the Fitting Factor by the match $\mathcal{M}$. An alternative variable, the mismatch ($\mathcal{M}\mathcal{M}$), is defined as $\mathcal{M}\mathcal{M}=1-\mathcal{M}$:
\begin{equation}
\mathcal{M}\mathcal{M} \equiv 1-\max_{\varphi,t}\frac{\langle h|h_{\rm UL} \rangle}{\sqrt{\langle h|h \rangle \langle h_{\rm UL}|h_{\rm UL} \rangle}},
\end{equation}
which is optimized over the GW phase $\varphi$ and the time of arrival $t$. The mismatch is independent on the amplitude (thus, on the distance of the source) and only quantifies the difference in shape between two waveforms. The inner product between two templates (A,B) is defined as 
\begin{equation}
\langle h_{\rm A}|h_{\rm B} \rangle = 2\int_{f_{\rm min}}^{f_{\rm max}} \frac{\tilde{h}_{\rm A}^*(f)\tilde{h}_{\rm B}(f)+\tilde{h}_{\rm A}(f)\tilde{h}_{\rm B}^*(f)}{S_{\rm n}(f)} df,
\end{equation}
where $S_{\rm n}$ is the one-sided noise power spectral density of the detector and * denotes complex conjugation.

A fiducial value for the detectability threshold can be estimated by using an approximation of the Bayes factor: 
$\ln B\sim(1-\mathcal{M}^2)\,\snr^2/2$ \cite{cornish-11}, where $B$ is the Bayes factor between the lensed and unlensed hypothesis. For detectability, the threshold on $\ln B$ can range from optimistic ($1$) to conservative ($10$) values respectively \cite{mishra-23-match}.
This expression simplifies to the Lindblom criterion \cite{lindblom-08,mcwilliams-10,gao-22,tambalo-23,savastano-23}\footnote{Following the discussions in \cite{caliskan-23,savastano-23}, the Lindblom criterion may be an optimistic approach because the deviation is assumed to be from lensing alone. One can take the parameter degeneracies into account by using the Fisher matrix formalism. However, the Fisher matrix approach has also some drawbacks \cite{vallisneri-08}: it can amplify numerical errors \cite{tambalo-23} and vary depending on the variables used. Here we use the Lindblom criterion for simplicity.} when $\mathcal{M}\mathcal{M}\ll1$, 
\begin{equation}
\ln B\sim\mathcal{M}\mathcal{M}\times \snr^2>(1-10),
\label{eq:detectability-criterion}
\end{equation}
where $\snr$ refers to the signal-to-noise ratio.
The (square of the) $\snr$ of a signal with strain $h$ is defined as 
\begin{equation}
\snr^2 \equiv 4
\int_{f_{\rm min}}^{f_{\rm max}}
\frac{|\tilde{h}(f)|^2}{S_{\rm n}(f)} df .
\label{eq:lensed-SNR-def}
\end{equation}
In LVK, for an $\snr\sim20$, the threshold mismatch for detectability is $\mathcal{M}\mathcal{M}\sim0.03$ ($3\%$) in the conservative case ($\ln B\sim 10$). In ET, the $\snr$ can be as large as $\sim 10^2-10^3$: for $\snr = 100$, 
the threshold mismatch in ET might be as small as $\mathcal{M}\mathcal{M}\sim10^{-4}$ ($\sim0.01\%$) and the condition for detectability would still hold in the optimistic case ($\ln B\sim 1$). However, it is more realistic to take the conservative $\ln B\sim10$ in Eq.~\eqref{eq:detectability-criterion}, which gives $\mathcal{M}\mathcal{M}\sim10^{-3}$ ($\sim0.1\%$). Since most ET signals would be at the limiting $\snr$ of $\sim 10$, the threshold mismatch would be similar to LVK, $\mathcal{M}\mathcal{M}\sim 0.01$ ($\sim 1\%$) in the optimistic case. 

\begin{figure}[t]
\centering
\includegraphics[width=\columnwidth]{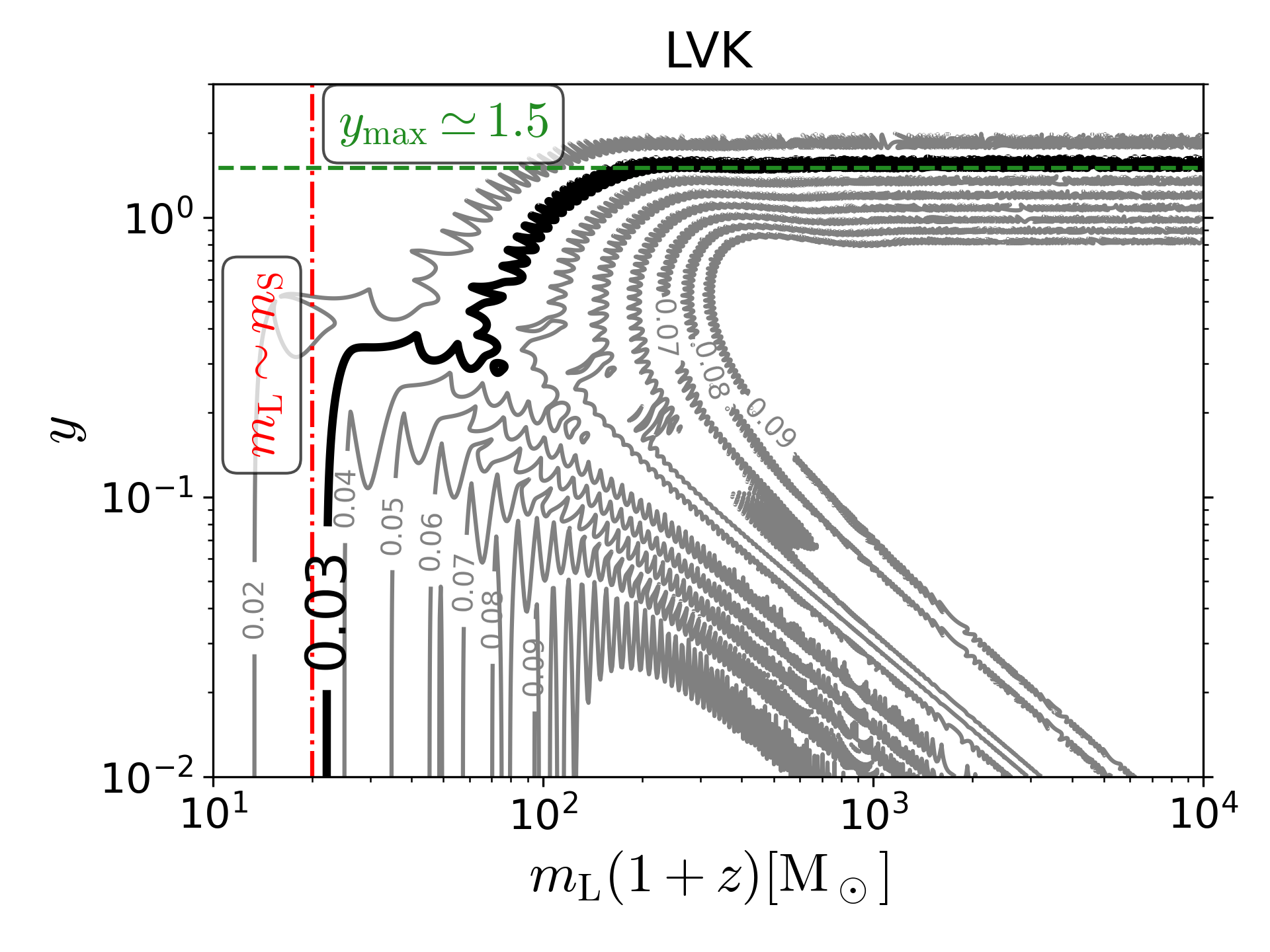}
\caption{Mismatch ($\mathcal{M}\mathcal{M}$) between a lensed waveform and the original waveform with the same source parameters, as a function of the redshifted lens mass $\ML(1+z)$ and the source displacement $y$. For this plot, we use a redshifted source mass of $\MS(1+z)=10 \, \msun +10 \, \msun$. At $\snr\sim20$, corresponding to the condition $\mathcal{M}\mathcal{M}>0.03$ in Eq.~\eqref{eq:detectability-criterion}, the distortions on the waveform due to lensing may be detected for $y\lesssim1.5$ and for $\ML/\MS \gtrsim 1$.}
\label{fig:appendix_match_O4}
\end{figure}
\begin{figure}[t]
\centering
\includegraphics[width=\columnwidth]{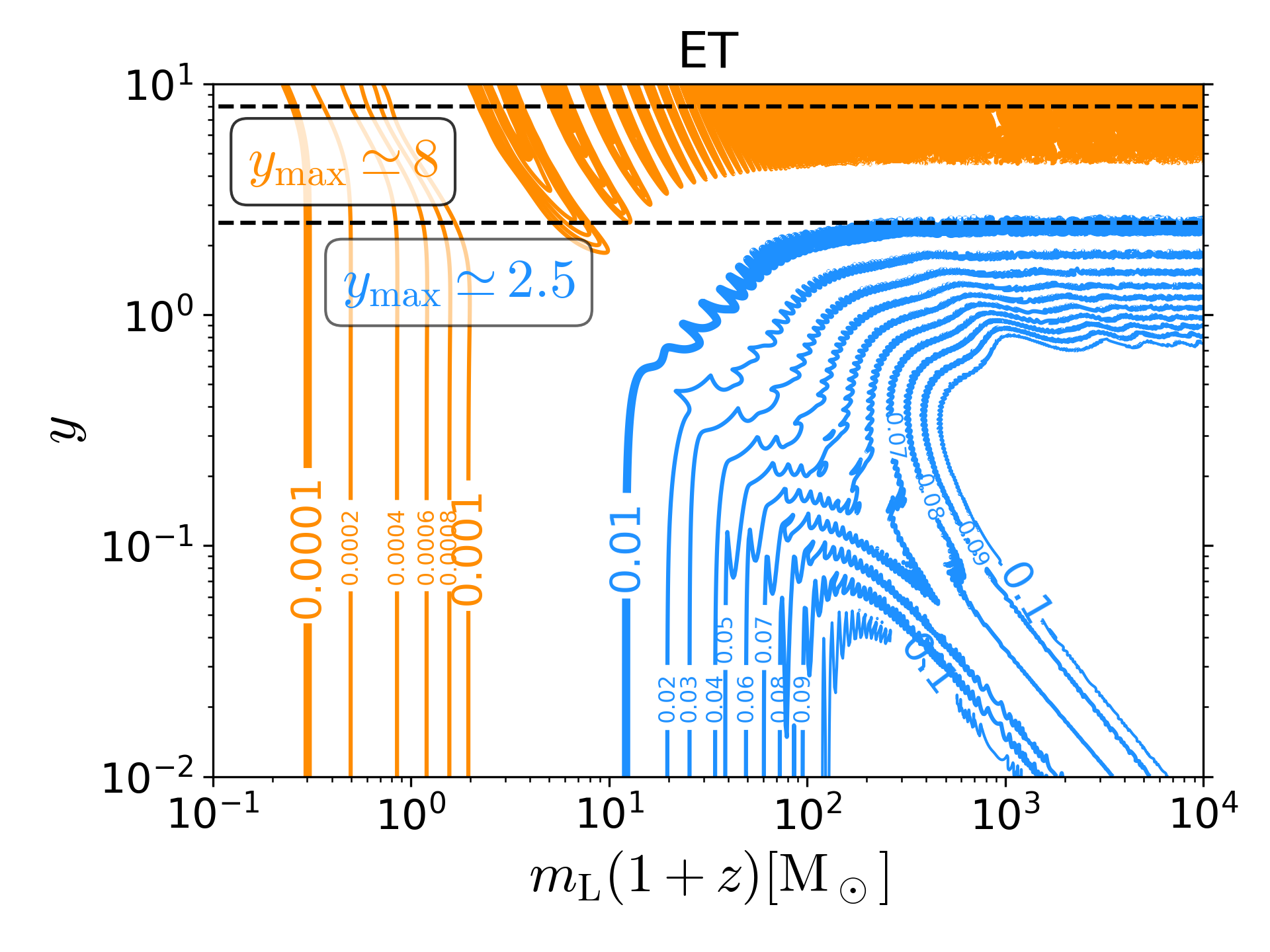}
\caption{Mismatch ($\mathcal{M}\mathcal{M}$) between a lensed waveform and the original waveform with the same source parameters, as a function of the redshifted lens mass $\ML(1+z)$ and the source displacement $y$. For this plot, we use a redshifted source mass of $\MS(1+z)=10 \, \msun +10 \, \msun$. For an $\snr\sim 10$, in blue, the optimistic value given by Eq.~\eqref{eq:detectability-criterion} $\mathcal{M}\mathcal{M}\sim 0.01$ ($\ln B\sim 1$) leads to threshold values at \mbox{$\ML/\MS\gtrsim0.6$}, \mbox{$\ymax\simeq2.5$}, while 
\mbox{$\mathcal{M}\mathcal{M}\sim 0.1$} corresponds to the conservative value $\ln B\sim 10$.
For $\snr\sim 100$, in orange, the threshold values may reach \mbox{$\ML/\MS\gtrsim0.02$}, \mbox{$\ymax\simeq8$} in optimistic conditions ($\mathcal{M}\mathcal{M}\sim 10^{-4}$) if no other distortions were present apart from lensing. However, it is more realistic to take a conservative value $\mathcal{M}\mathcal{M}\sim 10^{-3}$, which gives \mbox{$\ML/\MS\gtrsim0.1$}, \mbox{$\ymax\simeq5$}.}
\label{fig:appendix_match_ET}
\end{figure}

We explore the mismatch in the $(y, \ML(1+z))$ parameter space, similarly to \cite{RB-HU-OB-AL-2023,mishra-23-match}, using the \texttt{PyCBC} software. We assume the source parameters (component masses, inclination, orbital phase, polarization angle, distance and source position in the sky) are the same for both the lensed and unlensed templates. We use the \texttt{IMRPhenomD} waveform model, non-spinning and non-precessing, and the noise curves \texttt{aLIGOAdVO4T1800545} and \texttt{EinsteinTelescopeP1600143} for LVK and ET respectively.

Here we work with low redshift, $z\rightarrow0$. For $z>0$, both lens and source masses scale equally with $z$ because they have the same redshift. Therefore, the detectability is 
the same when we work with the redshifted masses.

We obtain the limiting values $\ymax$ and the minimum detectable lens mass $[\ML(1+z)]_{\rm min}$ from the mismatch thresholds. For the LVK O4 run, we take the threshold delimited by the line of $\mathcal{M}\mathcal{M}=0.03$ for $\snr\sim20$ shown in Fig.~\ref{fig:appendix_match_O4}. The space it encloses can be approximated as a rectangle with $\ymax$ as an upper boundary and $[\ML(1+z)]_{\rm min}$ as the left boundary. 
Lensing may be detectable for $y<\ymax$, $\ML(1+z)>[\ML(1+z)]_{\rm min}$.
We have analyzed the mismatch contour plot changing the source mass in the range $\MS(1+z) \in (20-200)\,\msun$ (assuming equal-mass binary components) to check if the detectability thresholds depend on $\MS(1+z)$. We find that the criterion for detectability depending on the masses for LVK is \footnote{We note that the ratio between the redshifted masses coincides with the ratio between the intrinsic masses, $\ML(1+z)/[\MS(1+z)]=\ML/\MS$, because the redshift is the same for the lens and the source.} 
\begin{equation}
\left(\ML/\MS\right)_{\rm LVK}\gtrsim 1, 
\label{eq:ml-ms-ratio-LVK}
\end{equation}
while $\ymax \simeq 1.5$ across the $\MS(1+z)$ range.

For ET (Fig.~\ref{fig:appendix_match_ET}), the values of the $\snr$ close to the threshold $\snr\sim 10$ may have a detectable lensing imprint for $\ML/\MS\gtrsim0.6$ and $\ymax \simeq 2.5$, similar to LVK results. For high $\snr\sim 100$, the threshold may be at low $\ML/\MS\gtrsim0.02$ and $\ymax \simeq 8$, although the real threshold is likely much more sensitive to other distortions. 

Since the $\snr$ distribution \cite[e.g.,][]{pieroni-22,iacovelli-22} above $\snr\gtrsim10$ decreases roughly as $\sim \snr^{-3}$, the high $\snr$ events are less likely to be observed than the low $\snr$ ones. In this study, we have not considered the modification of the $\snr$ distribution due to the lensing effect in the statistical sense. For wave optics lensing, it is not always the case that the observed $\snr$ increases due to lensing (as it does happen for strong lensing with a constant magnification): the observed $\snr$ can even decrease when not using the appropriate lensed template \cite{chan-25}. 

It is necessary to note that these thresholds only consider the lensing effect alone, to see where it would be significant enough to be detectable. In practice, additional effects such as precession, eccentricity, Earth rotation, etc. may create distortions that are degenerate with the lensing effect, and might hinder its identification. To obtain a more precise threshold, one may consider doing a Fisher matrix analysis or using waveforms that include precession, eccentricity and other effects.

Figure \ref{fig:detectability-imprints} summarizes the detectability thresholds as a function of the physical properties 
of the lensing system,
for the different environments considered in this work. The figure shows: 

\begin{figure*}[t]
\centering
\includegraphics[width=\columnwidth]{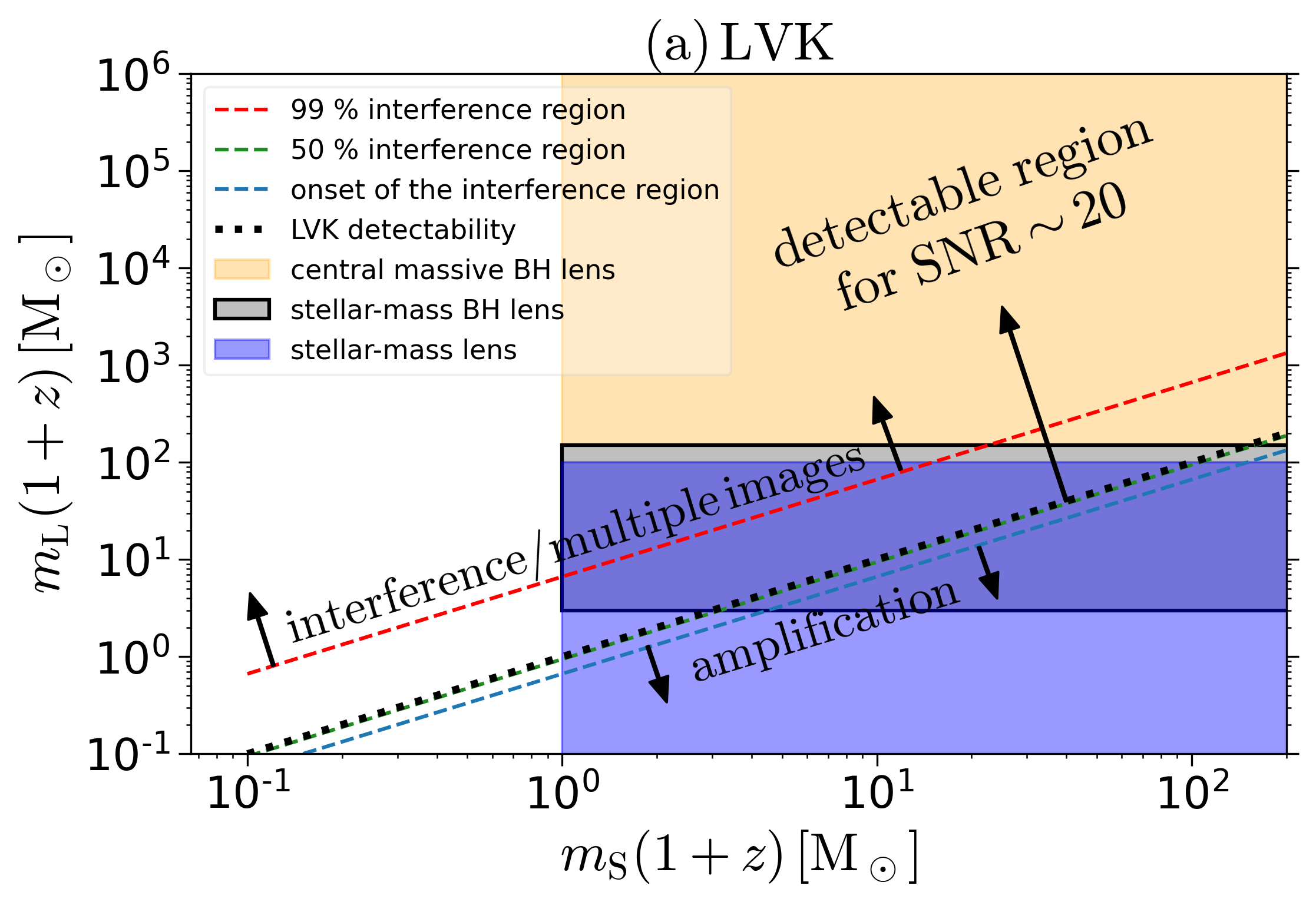}
\includegraphics[width=\columnwidth]{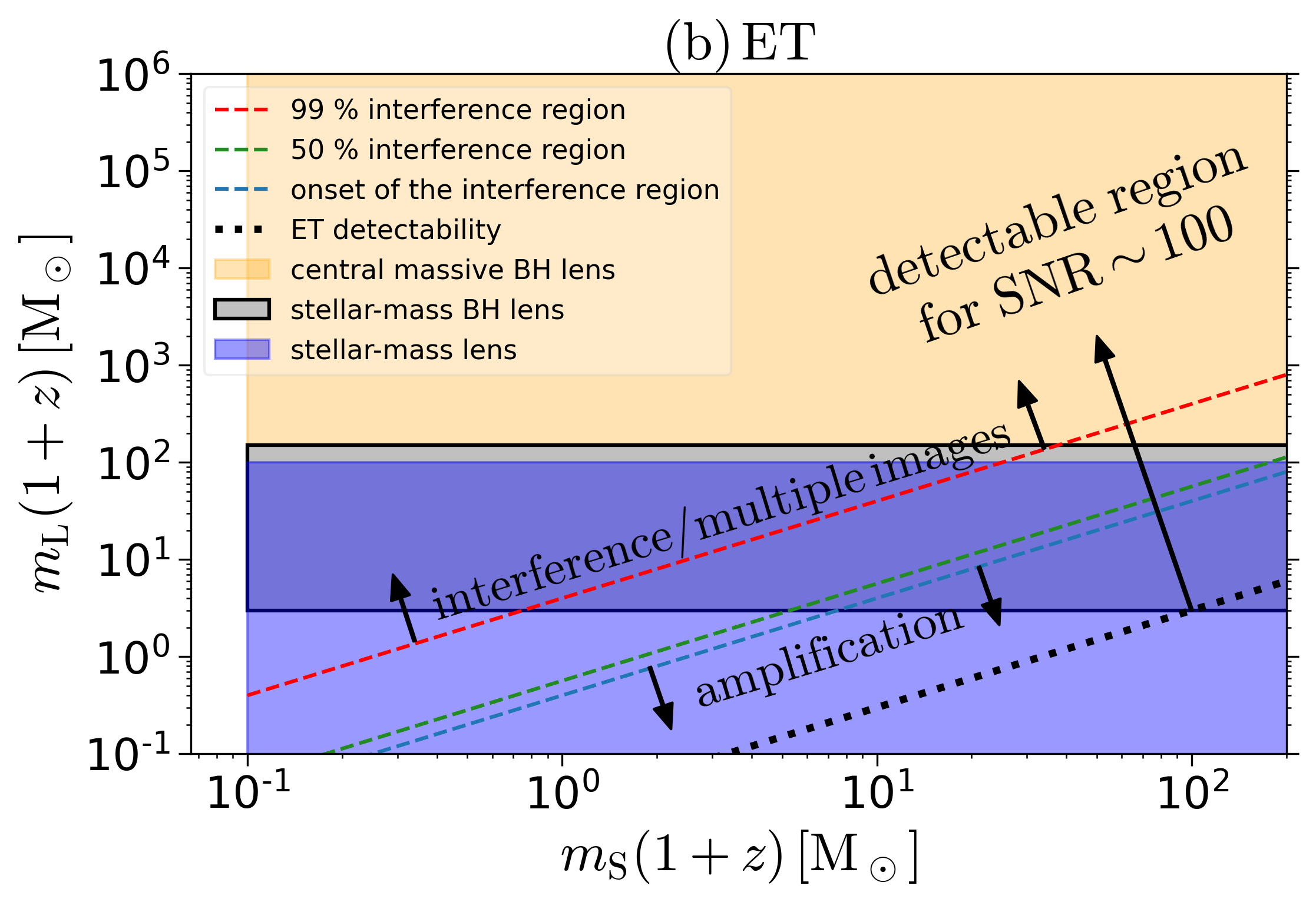}
\caption{Redshifted masses of the source and the lens in different self-lensing environments. The colored boxes indicate the type of lenses one would expect to find for each $\ML$ (fixing $z\simeq0$ to obtain these regions). The left and right boundaries of the boxes correspond to the GW detectors limitations to detect $\MS(1+z)$: LVK (a) can observe $\MS(1+z)\sim (1-200)\,\msun$, while ET (b), $\MS(1+z)\sim (0.1-200)\,\msun$. \textbf{Detectability}: self-lensing effects will be detected in the region above the detectability threshold, independently on $y$ (as long as $y<\ymax$). This threshold, assuming always an $\snr>8$, is shown in a black dotted line for (a) LVK for $\snr\sim20$ in the conservative case; (b) ET for $\snr\sim100$ in the optimistic case.  
\textbf{Imprint}: In general, 
there is only amplification when $\ML<\MS/\ymax$. When $\ML>\MS/\ymax$ (as long as $\yGO<\ymax$, using Eq.~\eqref{eq:yGO}), there is a region of interference. As $\ML$ increases, the probability of having lensing from the interference region increases in detriment of the amplification region (given by Eq.~\eqref{eq:prob-amp-only}). The dashed colored lines show the percentage covered by the interference region for (a) $\ymax=1.5$ and (b) $\ymax=2.5$. Below the onset ($p_{\rm diffr}=1$), there is amplification-only imprint. Above the $99\%$ line ($p_{\rm diffr}=0.01$), it can be considered that there is just interference, and well above that, multiple images can be resolved.}
\label{fig:detectability-imprints}
\end{figure*}

\begin{itemize}
    \item  The range of masses of the lens $\ML$ and the source $\MS$ for each environment, in colored boxes.

    \item The detectability thresholds: the lensing effect will be detectable above the black dotted line, corresponding to \mbox{$\ML>\MS$} for $\snr\sim20$ in LVK in the conservative case, and \mbox{$\ML>0.02\,\MS$} for $\snr\sim100$ in ET in the optimistic case, obtained in Sec.~\ref{sec:detectability} for Fig.~\ref{fig:detectability-imprints} (a) and (b) respectively. 
        
    Self-lensing 
    by a central massive BH lens (orange box)
    is well inside the detectability limit of both experiments. Self-lensing 
    by a stellar-mass lens, 
    by contrast, is fainter: 
    there is the possibility to observe a fraction of them, those at high $\snr$. Subsection \ref{subsec:imprint-environments} enters into the details of each environment.

    \item The lensing imprint regions 
(diffraction, interference, strong lensing) in dashed colored lines.
They depend on the value of $y$, taking as a reference the interference onset at \mbox{$\yGO\simeq\MS/\ML$} 
(Eq.~\eqref{eq:yGO}). 
Most detectable events will have an interference pattern or separate images, while chromatically amplified events are mostly undetectable, except for outlier events with very large $\snr$.
\end{itemize}

\subsection{Detectability of strong lensing}
\label{sec:detectability-strong}

For strong lensing (multiple images), both images should be detected above the threshold $\snr$ to claim lensing\footnote{Gravitational wave searches are currently based on thresholds set by false alarm rates rather than on the $\snr$s. However, the threshold $\snr$s mentioned here correspond to typical thresholds associated with specific false alarm rates for the corresponding detector networks, and can be taken as approximate thresholds.}.
The $\snr$ of the lensed image, $\snr_{\rm L}$, is related to the unlensed $\snr$ through the magnification of each lensed image, $\mu_+$ and $\mu_-$ for the brightest and faintest image, respectively (Eq.~\eqref{eq:magnif}),
\begin{equation}
\left(\snr_{\rm L}\right)^2 = 
4\int
\frac{|\tilde{h}_{\rm UL}(f)|^2|F|^2}{S_{\rm n}(f)} df = 
(\snr)^2 \mu_\pm\,,
\label{eq:lensed-SNR-def}
\end{equation}
where $|F|^2= \mu_\pm$ for each image in strong lensing.  
The source position $y$ is related to $\mu_+/\mu_-$ as \cite{schneider-92}
\begin{equation}
y = \left(\frac{\mu_+}{\mu_-}\right)^{1/4}
-\left(\frac{\mu_+}{\mu_-}\right)^{-1/4} ~.
\label{eq:ymuratio}
\end{equation} 
$\ymax$ depends on the maximum magnification ratio $\mu_+/\mu_-$, which we express in terms of the $\snr$ of each image using 
Eq.~\eqref{eq:lensed-SNR-def}, 
\begin{equation}
\left(\frac{\mu_+}{\mu_-}\right) = 
\left(\frac{\snr^+_{\rm L}}{\snr^-_{\rm L}}\right)^2 ~.
\end{equation}
We fix $\snr^-_{\rm L}$ to the minimum detectable $\snr$: $\left(\snr_{\rm L}^-\right)^{\rm min}=8$ for LVK \cite[e.g.][]{petrov-22},
and $\left(\snr_{\rm L}^-\right)^{\rm min} = 10$ for ET \cite{pieroni-22}. 
$\left(\snr_{\rm L}^+\right)^{\rm max}$ can be arbitrarily large, we need to take a reasonable value \cite{iacovelli-22}. For LVK, $\left(\snr_{\rm L}^+\right)^{\rm max}=50$, $\ymax\simeq 2$. For ET, $\left(\snr_{\rm L}^+\right)^{\rm max}=1000$, $\ymax\simeq 10$. Multiple imaging in strong lensing of GWs would then be detectable below these impact parameters.

\subsection{Characteristic imprints in astrophysical environments}
\label{subsec:imprint-environments}

In this subsection, we discuss the lensing imprints that can help distinguish the cases of self-lensing in a star cluster or in an AGN disk by a massive BH, from other cases of lensing. In Fig.~\ref{fig:flowchart_conclusions}, we summarize all the combinations in different environments, accounting for the different characteristic signatures (lensing imprints, eccentricity and characteristic polarizations), which we will use later to draw the conclusions. 

A general first distinction relates to the presence of diffraction and interference effects versus separate multiple images. If $\yGO > \ymax$, or equivalently $\ML/\MS < \ymax$, only the chromatic amplification imprint caused by diffraction is imprinted. When $\ML/\MS$ and $\ymax$ are comparable, interference patterns may be observed in addition, depending on $y$. For $\yGO \ll \ymax$, i.e. $\ML/\MS \gg \ymax$, self-lensing results in separate images. 

Microlensing by other objects in the line of sight may cause some imprints in the lensed signals when the lens-to-source distance $d_{\rm LS}$ is cosmologically large (foreground lenses). When the source is lensed by a galaxy, stellar mass objects acting as microlenses can have some effects even in the regime $\ML \ll \MS$ \cite{diego-19,mishra-21,shan-25}. In particular, saddle point images with high magnification can be most affected by microlensing \cite{shan-25}. In cases of self-lensing, microlenses should be much less important because of the low optical depths to self-lensing in these systems. Even in the central parts of AGNs, where the optical depth of the central BH can be large, the mass in stellar mass objects around the AGN in the lensing region is always much smaller than the central BH mass, so microlensing is not expected to be significantly present in self-lensing cases.

\subsubsection{Star clusters}
In star clusters, the mass of both BBH sources and potential stellar-mass lenses is of the same order, leading to $2\ML \sim \MS$, i.e. $\yGO\sim 2$.
The imprint is therefore in the diffraction and interference regimes.
The fraction of the cross section where diffraction (chromatic amplification) is present is
\begin{equation}
p_{\rm diffr} = \frac{\pi \yGO^2}{\pi \ymax^2} \simeq \frac{1}{\ymax^2} \left( \frac{\MS}{\ML} \right)^2,
\label{eq:prob-amp-only}
\end{equation}
which corresponds to the probability of having each imprint, depending on $y$ of a given signal. It is
shown for different values in Figs.~\ref{fig:detectability-imprints} (a) and (b) with dashed colored lines.
Both for LVK ($\ymax\simeq 1.5$) and ET ($\ymax\simeq 2.5$), the imprint is mainly diffraction because $\ymax\sim \yGO$.

Regarding their detectability, Fig.~\ref{fig:detectability-imprints} shows that only a fraction of the self-lensed sources in star clusters is detectable (above the black dotted line): 
those with a large $\ML/\MS$ value. 
For a random cluster lens and a
dynamical BBH merger, the mass ratio \mbox{$\ML/\MS$} is usually small and lies below the detectability line in LVK, while for ET a few of the signals with chromatic amplification with high $\snr$ may be detectable. A clear detection is possible only with \mbox{$\ML/\MS\gtrsim1/2$},
which in a GC implies
that the BBH merger source is not the central dynamical binary, but rather an off-center binary most likely formed via stellar evolution. 

When a central massive BH in a star cluster acts as lens, the large $\ML$ results in the imprint of an interference pattern or multiple images.
Separate images are detectable as long as both lie above the threshold $\snr$, as explained in Sec.~\ref{sec:detectability-strong}. 

\subsubsection{AGN disks}
\label{subsec:imprint-AGN}

The AGN case is a specific environment 
where there is always a central massive BH ---with a much higher mass than the source--- that can act as a lens. 
The detectable imprints can be the interference pattern and multiple images, depending on the mass of the lens. 

A unique signature may be present for self-lensing inside the AGN disk: the lensed signals 
should be seen with predominantly linearly polarized GW radiation, corresponding to $h_+$ for a directly overhead sky location of the source. The reason is that the AGN disk needs to be edge-on to allow us to see self-lensing by the SMBH or other BHs embedded in the disk. 
The orbital plane of the BBH merger source is thought to be preferentially aligned with the AGN disk \cite{king-05}, therefore it will also be edge-on with respect to us. Edge-on BBH merger sources emit pure $h_+$ signals. As a result, the detection of two linearly polarized signals would be a characteristic signal of self-lensing in an AGN disk.

\begin{figure*}
    \centering
    \includegraphics[width=\linewidth]{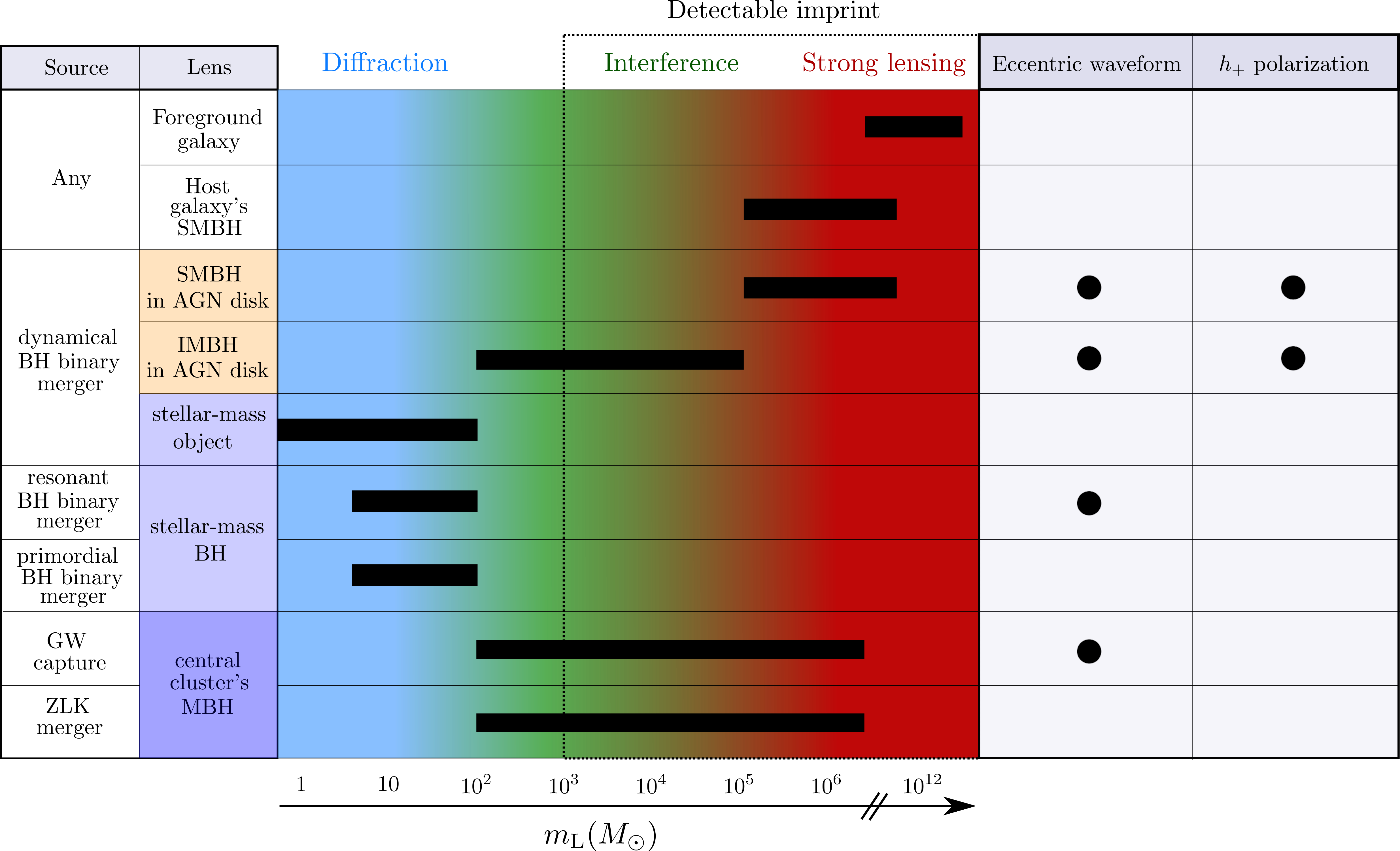}
    \caption{Summary of the characteristic signatures (lensing imprints, eccentricity and linear polarization) that we can expect to see for different combinations of sources and lenses in different environments. The color in the lens boxes corresponds to the environments: star clusters (blue) and AGN disks (orange), while the first two rows correspond to foreground lensing. We plot the range where the lensing imprints would occur for each case, depending on the lens mass $\ML$. For reference, IMBHs have a mass $\ML\sim(10^2-10^5)\msun$, SMBHs in NCs $\ML\sim(10^5-10^7)\msun$, and SMBHs in AGN disks (quasars) $\ML\sim(10^5-10^9)\msun$. Note that diffraction is generally not a detectable signature, while interference and multiple images are. Most cases have a unique combination of signatures that distinguishes them from others. 
    }
    \label{fig:flowchart_conclusions}
\end{figure*}

\section{Discussion}
\label{sec:discussion}

\subsection{Distinguishing self-lensing from foreground lensing}

Self-lensed signals may be confused with signals that are lensed by a foreground object between the host environment of the source and the observer. Is there a way of verifying that a signal is unequivocally self-lensed? We suggest several tests. 

\subsubsection{Unique environmental signatures of self-lensing}
Distinct self-lensing signatures can come either from the influence of the environment or from the geometrical configuration. 

While in a star cluster environment there are usually no additional environmental features to self-lensing, in an AGN disk it is possible to see some possible features, which we summarize here:

\textbf{Polarization of the images.} In the AGN disk case, we have argued that the disk's edge-on orientation provides a characteristic signature: the two strongly lensed images come with the same polarization, linearly polarized, which corresponds to $h_+$ for a source coming perpendicular to the GW detector arms. In contrast, a foreground strongly lensed event for the same detector geometry would have a mixed polarization of $h_\times$ and $h_+$, whose precise combination depends on the inclination\footnote{In standard General Relativity. Modified theories of gravity predict extra polarization modes.}. 

\textbf{Electromagnetic counterpart.} 
In an AGN disk, the presence of gas during the merger can produce an electromagnetic (EM) counterpart \cite{mckernan-19}. 

It is more likely to observe an EM counterpart for a face-on disk orientation \cite{graham-23}. However, the detection of self-lensing is only possible with a practically edge-on disk orientation, which makes it unfavourable to observe an EM counterpart. 
If we did observe a strongly lensed event with an EM counterpart from a thin AGN disk, the multiple image imprint would most likely be caused by foreground lensing. 

\textbf{Combination with environmental formation signatures on the waveform.} 
During their formation, GWs can acquire signatures from their interaction with the environment. The combination of these signatures with self-lensing imprints might improve the determination of their origin.

\begin{itemize}
\item \textbf{Waveform distortions}. When BBHs interact with their environment during the emission of GWs, their waveform can be distorted (apart from lensing effects), e.g. \cite{Zwick-25-EEI,leong-23}. In gaseous environments such as the AGN disk, the interaction of the binary with the gas may modify the GW waveform, e.g. \cite{caneva-24}. However, 
waveform distortions by gas (depending on whether gas speeds up or delays the merger time) are still subject to modelling uncertainty. 
Another distortion can come from dynamical interactions between the BBH and other massive objects, which can imprint a Doppler phase shift on the waveform \cite{samsing-24a,hendriks-24,samsing-24b}.

\item \textbf{Eccentricity}\\
GW signals from some types of sources, such as GW captures in resonant interactions and in single-single encounters close to a massive BH, are more likely to be eccentric due to their dynamical nature. 
The eccentricity alone may not be sufficient to distinguish the environment of the BBH. Nevertheless, a combination of the eccentricity with self-lensing imprints, especially the interference pattern or multiple images, might help determining their origin. 

In the cases we consider, GW captures during resonant interactions can be highly eccentric, and are only detectable when the lens is a central massive BH. In that case, the waveform would have imprints both of a high eccentricity and self-lensing.

\end{itemize}

\subsubsection{Time delay}
\label{sec:time-delay}
The time delay depends on $\ML$ and $y$ as $\Delta t \propto \ML y$ (from Eq.~\eqref{eq:time-delay}). While we expect $y$ to be similarly distributed for both the foreground and the self-lensing scenarios, the lens masses are going to be differently distributed in each case. Therefore, the time delay is also going to be different, which might help us distinguishing foreground lensing from self-lensing. This test is limited to strongly lensed signals, thus for the self-lensing by an IMBH/SMBH.

Consider that self-lensing is done by massive BHs with masses of $\ML\sim (10^5-10^9)\,\msun$.
Their mass distribution for the AGN disk case has been analyzed in \cite{gondan-kocsis-22}. In contrast, strong lensing is done by foreground galaxies with masses of $\ML\sim (10^9-10^{12})\,\msun$\footnote{Although low mass galaxies exist, most foreground lensing will be imprinted by the most massive galaxies, so we assume the contribution of the least massive galaxies is negligible.}.
The masses of these two populations of lenses (foreground lenses and self-lenses) are distributed differently. It should therefore be possible to statistically recover the two populations from the distribution of time delays.
The threshold mass that we take to classify foreground lensing and self-lensing,  $\ML=10^9\,\msun$, corresponds to a time delay \mbox{$\Delta t  \simeq 4\times10^4\,{\rm s}\simeq2\,{\rm days}$} for LVK, using $y\sim 1$.

\subsubsection{Optical depth and distance dependence}

Another way of distinguishing between self-lensing and lensing by foreground galaxies is by considering the distribution of distances $d_{\rm L}$ to lensed events.

The probability of self-lensing is an intrinsic property of the environment. It is practically independent on the distance $d_{\rm L}$\footnote{$\ymax$ will only gradually change with the mismatch threshold through the $\snr$, which will decrease as the distance increases.}. 
In contrast, foreground lensing has a probability that increases with the distance, scaling with the volume of possible lenses, 
$\tau \propto d_{\rm L}^3$ \cite[e.g.,][]{saas-fee-lectures-06}. 

For strong lensing, a statistical distribution of the events as a function of distance would give two separate distributions of the self-lensing probability as a function of distance. 
Even if the distributions were partially overlapping, self-lensing would contribute most at closest distances, while foreground lensing would contribute most at largest distances. 

For low-mass lenses $\ML\lesssim 10^3\,\msun$, we assume that the abundance of intergalactic low-mass lenses is negligible in comparison to self-lensing. Foreground lensing by stellar-mass objects in our Milky Way galaxy generally has an optical depth of $\tau<10^{-7}$ \footnote{We take a reference value from EM observations \cite[e.g.,][]{macho-project-00,eros-lmc-07,ogle-lmc-24,ogle-smc-25} as an upper threshold, because there are no lensed GW observations yet.}.
The optical depth for GWs is lower than the one for EM lensing because the number of potentially detectable lenses is lower, as discussed in Sec.~\ref{sec:prob-sigma}. Therefore, we consider that the contribution of foreground lensing by lenses in our own galaxy as well as foreground lensing in the host galaxy is small, additionally to the difficulty to detect their imprint. One should only be careful when the GW event comes from directly across the galaxy bulge, where the optical depth is higher, $\tau<10^{-5}$ \cite[e.g.,][]{macho-bulge-05,moaII-16,ogle-bulge-19}.

\subsection{Binary lens}
\label{sec:binary}
In this article we are considering a point mass lens model. However, in some cases the object acting as a lens can be a binary system. Is the point mass lens model still valid?

When the lens is a binary,
the point mass lens approximation holds when the projected distance $d_{\rm proj}$ between the binary components\footnote{The physical distance $d$ needs to be projected onto the lens plane, perpendicular to the line of sight, obtaining $d_{\rm proj}$. We assume the thin lens approximation.} is $d_{\rm proj}\ll\re$. In the opposite limit, $d_{\rm proj}\gg\re$, the binary components can also be approximated as individual lenses with external shear \cite{dominik-99} (Chang-Refsdal lenses \cite{chang-refsdal-79,chang-refsdal-84}). The binary lens model has to be taken into account either when $d_{\rm proj}\sim \re$, or very close to and inside the caustics, where extra images can contribute. 
Apart from this specific scenario, 
it is safe to assume point mass lenses. For the cases where the binary lens has to be considered, the imprint of a binary lens in the diffraction regime is similar to the point mass lens \cite{takahashi-thesis}. Self-lensing by a binary lens is left for future work. 

\subsection{Assumptions and limitations}
\label{sec:assumptions}

There are some assumptions that we have have used throughout this work, we discuss their validity here.\\

\textbf{Location of the migration traps.} We have used the results of the AGN disk models that predict migration traps at some given distances, which can potentially host BBHs \cite{mckernan-12,bellovary-16,secunda-19}.  
In \cite{gangardt-24} it is shown that one of their models coincides with \cite{bellovary-16,secunda-19} in having migration traps at 
$R_{\rm trap}\simeq 22\, \rs$, while $R_{\rm trap}\sim 10^3 \rs$ may be considered as a reasonable conservative value for different models. 
A recent study \cite{peng-21} claims closer migration traps, at $R_{\rm trap}=10 R_{\rm L,c}$ (used later in \cite{leong-25}), while another study \cite{grishin-24} finds them further away, at $R_{\rm trap}\sim(10^3-10^5) R_{\rm L,c}$. Although we used the values in \cite{bellovary-16,secunda-19}, the scaling $\tau\propto 1/R_{\rm trap}$ allows to easily estimate the probability for other values.

\textbf{Orientation of the BBH orbit relative to the AGN disk.} In this work, we have assumed that the orbital plane of BBHs is preferentially aligned with the plane of the AGN disk. 
This assumption allows us to connect the orbital alignment with the linear polarization of the signals, using the latter as an additional support for self-lensing in an AGN disk.
This prediction depends on the uncertainty of the alignment of the BBH orbit with respect to the AGN disk. This uncertainty can depend on time and history of the quasar and its internal perturbations. The BBH would also be misaligned if it was precessing.

\textbf{Polarization.} When we state that only linear polarization $h_+$ implies an edge-on configuration, we consider the predominant $(2,2)$ mode in the GW signal. Higher modes also have only linear polarization in the edge-on emission. In fact, compared to other inclinations, it is in the edge-on configuration that these higher modes have a higher relevance with respect to the (2,2) mode. In some circumstances (such as for a precessing BBH), higher modes can be significant, and have contributions from both $h_\times$ and $h_+$ in the edge-on configuration. We do not consider precession in this work. 

Observational prospects for the polarization content are still limited in current detectors, due to degeneracies with the luminosity distance and the dependence on the inclination. 
Nonetheless, edge-on configurations have better prospects of being identified than other inclinations \cite{usman-19}. 

Another assumption we made regarding the polarization is that we neglected the effect of lensing in the rotation of the polarization plane, which would only be significant in the very strong field regime ($\dLS\sim\rs$) \cite{takahashi-03}. 

\textbf{Additional environmental effects.} We assumed a non-spinning lens. We neglect the gravitation spin Hall effect \cite{oancea-24}. We have neglected the Doppler boost and beaming effect, as well as the Shklovskii effect for moving GW sources, although they may be taken into account in AGN disks. 
Environmental effects and eccentricity should be included in the future to better estimate the detectability of realistic signals (for example, highly eccentric lensed waveforms).

\textbf{Detectability criterion.} For wave effects, we used the mismatch as a detectability criterion, while not considering the effect of the $\snr$ in the false alarm rate \cite{chan-25}. In order for the criterion in Eq.~\eqref{eq:detectability-criterion} to be accurate, it is necessary to assume a Gaussian and stationary noise, which in current LVK detections is not always the case. However, for the purpose of analyzing the detectability in general for different detectors and using it to get an estimate of $\ymax$, we consider the criterion used in this work to be a good enough estimate. 

\textbf{Diffraction by stellar fields inside the environment.}
In this work, we have neglected the weak lensing effect of surrounding objects inside the environment. In self-lensing, the lens and the source are very close to each other. Therefore, not many objects are close enough to the line of sight to contribute significantly to the external shear.

\subsection{Implications}

\textbf{Breaking the degeneracy on the origin of eccentric sources.} 
GW sources from dynamical channels (GW captures in star clusters and mergers in AGN disks) have residual eccentricities at LVK and ET frequencies. Eccentricity is  used as an indicator of the origin of the GW source. 
However, the eccentricity distributions in GW captures in different environments are expected to be similar. Self-lensing can break this degeneracy: (i) a strongly lensed eccentric signal with linear polarization is most likely to come from an AGN disk; (ii) a strongly lensed eccentric signal with arbitrary polarization is most likely to come from a single-single GW capture near the massive BH in the center of a NC. 

\textbf{Inclination of BBH orbital planes in AGN disks.}
The information from the self-lensing probability and the polarization, combined with an independent constraint to the orbital inclination of the BBHs (e.g., through the probability distribution, \cite{samsing-22}) might be used to constrain the relation between the alignment of the BBH orbit relative to the AGN disk, knowing that the AGN disk is edge-on when we observe self-lensing.

\section{Conclusions}
\label{sec:conclusions}

In this work we provide estimates for the probability of lensing of GW sources by nearby lenses (self-lensing) and quantify the expected lensing imprint, to assess whether self-lensing can be a discriminator for different formation channels of BBH mergers. A summary of the probability of detectable self-lensing in environments with a massive central BH is given in Fig.~\ref{fig:summary-probabilities} and a summary of the characteristic signatures of the different environments is shown in Fig.~\ref{fig:flowchart_conclusions}. 
 
We conclude that:
\begin{itemize}
    \item Self-lensing by stellar-mass BHs in star clusters is unlikely to be detected unless the $\snr$ is very large.
    First, because it is improbable: the optical depth depends on the velocity dispersion as $\tau\sim \sigma^2/c^2\simeq10^{-7}$. 
    Even if we take into account the effect of nearby BHs in resonant interactions, we find $\tau<10^{-5}$. 
    Second, most self-lensed signals for dynamically formed BBH have lenses that are less massive than the source masses ($2\ML\lesssim\MS$), because a dynamical BBH tends to form from the most massive BHs in the system. These are mostly undetectable in LVK as shown in Fig.~\ref{fig:detectability-imprints}, unless the $\snr$ is very large. 
    Even in ET, with which lower values of $\ML/\MS$ may be detected with higher $\snr$ events, it is impractical to detect wave optics self-lensing, due to the low fraction of high $\snr$ signals and the difficulty of disentangling self-lensing distortions from other effects.
    \item 
    We quantify the probability through $\tau$ for self-lensing of single-single GW captures and ZLK induced mergers in the Bahcall-Wolf cusp around a massive BH in NCs. 
    
    The probability of single-single GW captures increases with lens mass, where 
    $\tau\sim 4\times 10^{-5}\, (m_{\rm L, c}/[10^6\,\msun])^{0.97}$ when $m_{\rm L, c}>3\times10^5\,\msun$, for $\sigma=30\,\kms\,(m_{\rm L, c}/[10^6\msun])^{0.2}$, $\MS=20\,\msun$, $M_{\rm clus}=10\,m_{\rm L, c}$, $\ymax=1$. 
    
    ZLK mergers have a probability comparable to single-single captures at high masses, while at low lens masses ($\ML<3\times10^5\,\msun$) the probability is higher than the one of single-single GW captures. The self-lensing probability for ZLK mergers is given by $\tau\simeq 3\times10^{-5} (m_{\rm L,c}/[10^6 \,\msun])^{2/3} \,\ymax^2$.
    In these two cases, the GW signal is easily detectable as two strongly lensed images, as long as both exceed the threshold $\snr\gtrsim8$. 
    
    We expect to detect a few (\mbox{$\sim 1 - 10$}) self-lensed events from single-single GW captures and/or ZLK merger sources\footnote{In the case where all the mergers come from one of these channels. Otherwise, the relative contributions from different channels need to be considered.} in next generation GW detectors such as ET and CE, given the high number of expected GW detections, \mbox{$\sim (10^5-10^6)\, {\rm events}/{\rm year}$} \cite{maggiore-20-ET-case,evans-21-CE-white-paper}.
    \item An AGN disk is the most probable environment to have self-lensing, both for the case of central lensing by the SMBH and for IMBHs lenses embedded in the disk. For the central SMBH lensing a BBH merger in a migration trap at a distance of $25$ Schwarzschild radii from the SMBH, $\tau \simeq 0.02$ (consistent with previous studies \cite{gondan-kocsis-22}). This case is also easily detectable as two separate images. 
    For an embedded IMBH lens, \mbox{$\tau \simeq 2 \times 10^{-5} \,(m_{\rm L, lat}/[10^3\, \msun]) \,([10^6\, \msun]/m_{\rm c})\,\ymax^2$}, where the lensing imprint would be an interference pattern of the emerging images. A unique imprint in this environment is the linear polarization $h_+$, arising from the alignment of the orbit of the binary with the AGN disk plane. This characteristic signature enables us to distinguish the signal as coming specifically from an AGN disk.
    \item It is possible to qualitatively distinguish the environment when the signal is strongly lensed, as indicated in Fig.~\ref{fig:flowchart_conclusions}. First, the value of the time delay can help us classify foreground lensing and self-lensing as discussed in Sec.~\ref{sec:time-delay}. Then, the polarization of the images can allow us to recognize the case of self-lensing in the AGN disk (linear $h_+$ polarization). 
    On the other hand, lensed images with time delays $\Delta t\lesssim 2\,{\rm days}$ with a 
    combination of $h_+$ and $h_\times$ polarizations will come from star clusters, when there is lensing by the central massive BH. In star clusters the dominant self-lensed sources are single-single GW captures close to massive SMBHs, likely to be eccentric, and ZLK mergers. The higher likelihood of eccentricity in single-single GW captures compared to ZLK mergers can support their identification. 
\end{itemize}

\section*{Acknowledgements}
The authors are grateful to Anuj Mishra for the insightful discussion on LVK detectability criteria, for carefully reviewing the manuscript and for providing useful suggestions. We appreciate the anonymous referee's comments and suggestions, which helped clarify and improve the text. The authors would like to acknowledge Andrew Lundgren for the availability of the code that was used to compute the mismatch. We are grateful for discussions with Fabio Antonini, Daniel Mar\'{i}n Pina, Oleg Bulashenko, Tomas Andrade, Kazushi Iwasawa, Nadia Blagorodnova, Sara Rastello and the Virgo and Stellar Dynamics groups at ICCUB, and later discussions with Alessandro Trani, Juno Chan, Miguel Zumalac\'{a}rregui, Johan Samsing, Alexander Dittmann, Javier Roulet and Daniel D'Orazio. 
HU appreciates conversations with Miko\l aj Korzy\'{n}ski and Johan Samsing about transverse Doppler and phase shifts. We are grateful to Esther Pallar\'{e}s for logistical help.

We have used the \texttt{PyCBC} software\footnote{\href{https://zenodo.org/records/10473621}{https://zenodo.org/records/10473621}} to compute the waveforms, the (mis)match and the $\snr$.

The authors acknowledge financial support from  PID2021-125485NB-C22 (HU, MG), PID2022-137268NB-C52 (JME), PID2024-159689NB-C22 (HU), PID2024-155720NB-I00 (MG) and CEX2019-000918-M, CEX2024-001451-M (HU, MG, JME) funded by 
MCIN/AEI/10.13039/501100011033. HU and MG also acknowledge SGR-2021-01069 (AGAUR, Generalitat de Catalunya). 
HU acknowledges financial support from the FI-SDUR 2023 predoctoral grant (AGAUR, Generalitat de Catalunya).

\section*{Data availability}

The code to produce the (mis)match data can be found in the GitHub repository \href{https://github.com/helenaubach/GW-Lensing-match}{https://github.com/helenaubach/GW-Lensing-match}.

\appendix

\section{General probability of self-lensing by a member of a star cluster}
\label{appendix:general-probability} 
The optical depth along the line of sight is defined as \cite{schneider-92, narayan-96}
\begin{align}
\tau \equiv
 \int_0^\infty& 2\pi R  dR
\int_{-\infty}^{\infty} d z_{\rm L} n_{\rm L}(R,z_{\rm L})& \nonumber\\
& \left\{ \int_{z_{\rm L}}^\infty d z_{\rm S} p_{\rm S}(R,z_{\rm S}) \Sigma_{\rm L}(z_{\rm S}, z_{\rm L})\right\} ~,
\label{eq:optical_depth_general}
\end{align}
where $R$ is the impact parameter from the cluster center and $z_{\rm L}$, $z_{\rm S}$ are the positions of lenses and sources respectively (not to confuse with the redshift), in the direction of the line of sight, with origin at $d_{\rm L}$. Here the lensing cross section is $\Sigma_{\rm L}=2\pi \ymax^2 \rs (z_{\rm S}-z_{\rm L})$ (Eq.~\eqref{eq:cross-section}), where we used $d_{\rm LS}= z_{\rm S} - z_{\rm L}$, and the maximum impact parameter of the source from the lens for detection is $\ymax$.
The number density of potential lenses is $n_{\rm L}$, and $p_{\rm S}$ is the probability of finding the source at some spatial position, defined such that $\int p_{\rm S}{\rm d}V=1$. For the special case of having the source at the center (describing dynamically-formed sources), $p_{\rm S}$ is given by Dirac delta distributions at the origin. When all lenses
are also potential sources (as is the case for optical lensing), $p_{\rm S}=n_{\rm L}(R, z_{\rm S})/N_{\rm L}$, where $N_{\rm L}$ is the number of lenses. 

To quantify the $\tau \propto (\sigma/c)^2$ relation, we assume that the number density of lenses is given by a Plummer model \cite{1911MNRAS..71..460P}
\begin{equation}
n(r)=\displaystyle{\frac{3N_{\rm L}}{4\pi r_0^3}  \left( 1 + \frac{r^2}{r_0^2} \right)^{-5/2}},
\end{equation}  
where $r$ is the radial distance to the center and $r_0$ is the Plummer scale radius. 

The mean optical depth follows then from integrating 
Eq.~\eqref{eq:optical_depth_general} with $p_{\rm S}=n_{\rm L}/N_{\rm L}$: 
\begin{align}
\bar\tau&= \frac{\pi^2}{8} 
\frac{GN_{\rm L}\ML
}{r_{0} c^2} \ymax^2,\\
&= 4\pi\left(\frac{\sigma}{c}\right)^2 \ymax^2,
\label{eq:optical_depth_general_plummer}
\end{align}
where we used $G M_{\rm clus}/r_{0} = (32/\pi) \sigma^2$, with $\sigma$  the mean one-dimensional velocity dispersion of the 
Plummer model. For sources in the center of the cluster, $\tau$ is a factor of $\sim3$ lower than $\bar\tau$. 

We previously assumed that all objects in the cluster are potential lenses. 
However, not all the objects that give rise to the potential are detectable lenses. In a star cluster, the most massive BHs are the likely lenses: it is necessary that $\ML\gtrsim m_{\rm S}$ (Eq.~\eqref{eq:ml-ms-ratio-LVK}) for LVK detectability. For ET, the detectability threshold may allow us to see lower mass lenses for higher values of the $\snr$, as long as we could distinguish lensing from other waveform variations.

To write the detectable optical depth $\bar\tau_{\rm det}$ in terms of $\sigma$ of the cluster, we use the mass fraction 
of the detectable lenses relative to the cluster mass 
$f_{\rm L}=M_{\rm Lenses}/M_{\rm clus}$, 
and the (scale) radius of the detectable lens population relative to the scale radius of the cluster ($f_{r_0} = r_{0, \rm Lenses}/r_{0, {\rm cl}}$) and assume that the BH system is self gravitating to write Eq.~\eqref{eq:optical_depth_general_plummer} as
\begin{align}
\bar\tau_{\rm det} & = 4\pi \frac{f_{\rm L}}{f_{r_0}}\left(\frac{\sigma}{c}\right)^2 \ymax^2\\
&\simeq10^{-7} \, 
\frac{f_{\rm L}}{0.05}\,\frac{0.075}{f_{r_0}} 
\left(\frac{\sigma}{30\,\kms}\right)^2 \ymax^2.
\label{eq:optical_depth_general_plummer_BH_appendix}
\end{align}
We scaled the result to a BH mass fraction of 5\%, typical for a canonical stellar initial mass function and a high BH retention fraction \cite{breen-heggie-13}.

\section{Self-lensing probability for single-single GW captures near a  central massive BH}
\label{appendix:gw-capture}

Here we derive the optical depth of single-single GW captures occurring in a Bahcall-Wolf cusp \cite{1976ApJ...209..214B} around a central massive BH, lensed by the central massive BH. The rate of single-single GW captures in this scenario was studied with Fokker-Planck models \cite{oleary-09}. Here we derive simple analytic relations that reproduce those results to be able to derive the optical depth. 
We express the normalized probability that a single-single GW capture occurs at a radius $r$ from a central massive BH as
\begin{equation}
p_{\rm S} = \frac{\Gamma_{\rm capt}(r)}{\Gamma_{\rm tot}}.
\label{eq:prob-gw-capt-radius}
\end{equation}
Here $\Gamma_{\rm capt}(r)$ is the single-single GW capture rate per unit volume at $r$, which is given by
\begin{equation}
\Gamma_{\rm capt}(r) \equiv n_{\rm BH}(r) \Gamma_{\rm ss}(r),
\label{eq:gw-capt-radius}
\end{equation}
where $n_{\rm BH}(r)$ is the number density profile of BHs and $\Gamma_{\rm ss}(r)$ is the capture rate of a single BH at $r$. 
The normalization $\Gamma_{\rm tot}$ is the total capture rate in the volume within $r_{\rm min}<r<r_{\rm max}$
\begin{equation}
\Gamma_{\rm tot} = \int_{r_{\rm min}}^{r_{\rm max}} \Gamma_{\rm capt}(r) 4\pi r^2 dr.
\label{eq:gw-capt-tot}
\end{equation}
For $n_{\rm BH}(r)$ we assume a Bahcall-Wolf cusp, which has a (mass) density profile of the form $\rho_{\rm cusp}(r)\propto r^{-7/4}$, such that
\begin{align}
 n_{\rm BH}(r) &= \frac{\rho_{\rm cusp}(r)}{m_{\rm BH}}  \label{eq:rho-cusp1}\\
 &\simeq \frac{\rho_{\rm h}}{m_{\rm BH}}
\left( \frac{m_{\rm L,c}}{M_{\rm clus}} \right)^{1/4} \left(\frac{r}{r_{\rm h}}\right)^{-7/4}\label{eq:rho-cusp2} \\
 &\equiv n_{\rm h}\left(\frac{r}{r_{\rm h}}\right)^{-7/4}.  \nonumber
\end{align}
Here $\rho_{\rm h}=3M_{\rm clus}/(8\pi r_{\rm h}^3)$ is the density within the half-mass radius ($r_{\rm h}$), and $m_{\rm BH}$ is the mass of the individual stellar-mass BHs.
To find the constant of proportionality $n_{\rm h}$, we assumed that the cluster is relaxed so we can use the scaling relations of Heggie et al. \cite{2007PASJ...59L..11H}. 
Defining $r_{\rm cusp}$ as the radius where the cusp joins the core of the cluster, we can write: $\rho_{\rm cusp}(r) = \rho_{\rm c}(r/r_{\rm cusp})^{-7/4}$, where $\rho_{\rm c}$ is the density in the core. Heggie et al. show that 
\begin{equation}
r_{\rm cusp} = r_{\rm h}\frac{m_{\rm L,c}}{M_{\rm clus}}
\label{eq:rcusprh}
\end{equation}
and because the cluster is approximately isothermal between the core and $r_{\rm h}$, we can write $\rho_{\rm c} =\rho_{\rm h}(r_{\rm c}/r_{\rm h})^{-2}$, where $r_{\rm c}$ is the core radius. 
The final step to arrive at Eq.~\eqref{eq:rcusprh} is the relation between $r_{\rm c}$ and $r_{\rm h}$ for which Heggie et al. find  $r_{\rm c}/r_{\rm h} = (m_{\rm L,c}/M_{\rm clus})^{1/4}$. 

The rate of single-single GW captures at a given radius is \cite{samsing-20}
\begin{equation}
\Gamma_{\rm ss} \sim G R_{\rm S} \frac{m_{\rm BH}n_{\rm BH}(r)}{\sigma} \left( \frac{\sigma^2}{c^2} \right)^{-2/7},
\label{eq:gamma-ss}
\end{equation}
where $R_{\rm S}$ is the Schwarzschild radius of the individual stellar-mass BHs and $\sigma$ is the velocity dispersion. We do not need the constant of proportionality, because we are interested in the normalized probability (Eq.~\eqref{eq:prob-gw-capt-radius}).

Within the cusp, the potential is dominated by the central BH, such that the dispersion can be written as
\begin{equation}
\sigma \sim \left( \frac{Gm_{\rm L,c}}{r} \right)^{1/2}.
\end{equation}
Using this relation in Eq.~\eqref{eq:gamma-ss} and substituting $n_{\rm BH}(r)$ from Eq.~\eqref{eq:rho-cusp2} we find 
\begin{align}
&\Gamma_{\rm ss} \sim \frac{\sigma^3}{Gm_{\rm BH}}
\left( \frac{m_{\rm L,c}}{M_{\rm clus}}\right)^{1/4}
\left( \frac{Gm_{\rm BH}}{c^2r_{\rm h}}\right)^{5/7}
\left( \frac{r}{r_{\rm h}} \right)^{-27/28},\\
&\Gamma_{\rm ss}  \equiv  \, 
  \Gamma_{\rm h}\left( \frac{r}{r_{\rm h}} \right)^{-27/28}. 
  \label{eq:gamma-ss-norm}
\end{align}

The integral limits of Eq.~\eqref{eq:gw-capt-tot} are $r_{\rm min}=r_2$, $r_{\rm max}=r_{\rm cusp}$, where $r_{\rm cusp}$ is given by Eq.~\eqref{eq:rcusprh} and
$r_2$ corresponds to the radius that can contain at least 2 stellar-mass BHs of mass $m_{\rm BH}$. We find  $r_2$ from solving $\int_0^{r_2}n_{\rm BH}dV =2$ and by using Eq.~\eqref{eq:rho-cusp2}, we obtain
\begin{equation}
r_2 =\left(\frac{5}{3}\right)^{4/5}r_{\rm h} \left( \frac{m_{\rm BH}}{M_{\rm clus}} \right)^{3/5} \left( \frac{m_{\rm BH}}{m_{\rm L,c}} \right)^{1/5}.
\label{eq:r2}
\end{equation}

Using Eq.~\eqref{eq:gamma-ss-norm}, Eq.~\eqref{eq:gw-capt-radius} can be written as
\begin{equation}
\Gamma_{\rm capt} = n_{\rm h}\Gamma_{\rm h}\left(\frac{r}{r_{\rm h}}\right)^{-19/7}.
\label{eq:Gamma_capt}
\end{equation}
We can compare this radial dependence of $\Gamma_{\rm capt}$ to the Fokker-Planck results of \cite{oleary-09} (their Fig.~7). They find that $d\Gamma_{\rm capt}/d\log r$ is approximately constant, which agrees with our scaling of Eq.~\eqref{eq:Gamma_capt}, $d\Gamma_{\rm capt}/d\log r\propto r^{2/7}$. Also, for their adopted model parameters of $m_{\rm L,c} = 3.5\times10^6\,\msun$, $m_{\rm BH} = 25\,\msun$ and $\sigma=75\,\kms$, which we assume is due to $M_{\rm clus}=3\times10^7\,\msun$ and $r_{\rm h}\simeq3\,\pc$ we find $r_2\simeq10^{-4}\,\pc$, which agrees well with the rapid increase of the merger rate seen between $\sim(1-3)\times10^{-4}\,\pc$ shown in their Fig.~7. 
Combined with Eq.~\eqref{eq:gw-capt-tot} we can now compute the total capture rate needed for the normalization
\begin{align}
\Gamma_{\rm tot} &= 4\pi n_{\rm h}\Gamma_{\rm h}r_{\rm h}^2 \int_{r_2}^{r_{\rm cusp}} \left(\frac{r}{r_{\rm h}}\right)^{-5/7} dr \nonumber \\
&= 14\pi n_{\rm h}\Gamma_{\rm h}r_{\rm h}^3 \left[\left(\frac{r_{\rm cusp}}{r_{\rm h}}\right)^{2/7}-\left(\frac{r_2}{r_{\rm h}}\right)^{2/7}\right],
\end{align}
obtaining for Eq.~\eqref{eq:prob-gw-capt-radius}
\begin{equation}
p_{\rm S} = \frac{(r/r_{\rm cusp})^{-19/7}}{14\pi r_{\rm cusp}^{3}(1- (r_2/r_{\rm cusp})^{2/7})}. 
\label{eq:prob-gw-capt-radius-developed}
\end{equation}

To obtain the optical depth, we use Eq.~\eqref{eq:optical_depth_general} using $p_{\rm S}$ from Eq.~\eqref{eq:prob-gw-capt-radius-developed}, in the particular case of having the lens at $z_{\rm L}=0$, which implies $R=0$:
\begin{align}
\tau &= \int_{r_2}^{\rm r_{cusp}} d z_{\rm S} \, p_{\rm S}(z_{\rm S})\Sigma_{\rm L}(z_{\rm S})\label{eq:tau_app}\\
&=\frac{\rs \ymax^2}{7  (r_{\rm cusp}^{2/7}-r_2^{2/7})} \int_{r_2}^{\rm r_{cusp}} dz_{\rm S} \, z_{\rm S}^{-12/7}\\
&=\frac{\rs \ymax^2}{5}\,\frac{ r_2^{-5/7}-r_{\rm cusp}^{-5/7}}{r_{\rm cusp}^{2/7}-r_2^{2/7}},\label{GW-capture-full-result}\\
&\simeq \frac{\rs \ymax^2}{5r_{\rm cusp}}\left(\frac{r_2}{r_{\rm cusp}}\right)^{-5/7} ~, \label{GW-capture-approx-result}
\end{align}
where we used $\Sigma_{\rm L}=2\pi\ymax^2\rs z_{\rm S}$. In the final step we made the approximation $r_2\ll r_{\rm cusp}$. The expression can be further developed by substituting $r_2$ and $r_{\rm cusp}$. If we use $r_{\rm h}=GM_{\rm clus}/\sigma^2$, the final result is 
\begin{align}
\tau \simeq 4\times 10^{-5}\ &\left( \frac{\sigma}{30\,\kms}\right)^2
\left(\frac{10 \, m_{\rm L,c}}{M_{\rm clus}}\right)^{2/7} \times \nonumber \\
\times&
\left(\frac{m_{\rm L,c}}{10^6\,\msun}\right)^{4/7}
\left(\frac{m_{\rm S}}{20\,\msun}\right)^{-4/7}\, \ymax^2. 
\label{eq:prob-ratio-approx}
\end{align}

\begin{figure}[h]
\includegraphics[width=\columnwidth]{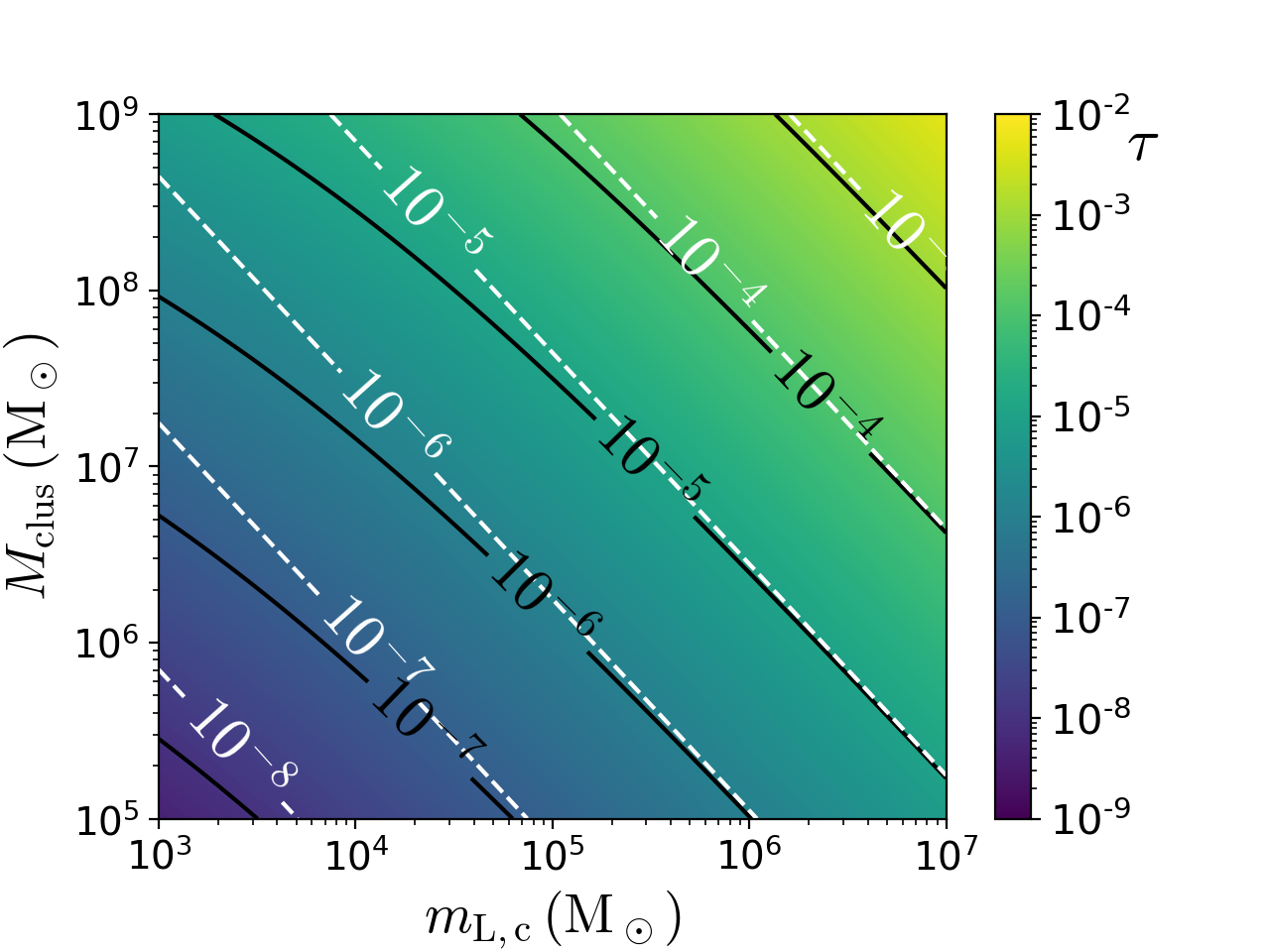}
\caption{Probability of self-lensing for GW single-single captures in a Bahcall-Wolf cusp
as a function of the mass of the central lens $m_{\rm L,c}$ and the mass of the cluster $M_{\rm clus}$, for $m_{\rm BH}=10\,\msun$. The black solid lines correspond to the complete solution given by Eq.~\eqref{GW-capture-full-result}, while the white dashed lines correspond to the approximation given by Eq.~\eqref{eq:prob-ratio-approx}.}
\label{fig:prob-ratio-M-MCBH}
\end{figure}

In Fig.~\ref{fig:prob-ratio-M-MCBH}, we compare the full result in Eq.~\eqref{GW-capture-full-result} in black solid lines with the approximation in Eq.~\eqref{eq:prob-ratio-approx} in white dashed lines. We can see that the approximation is valid for massive central BHs ($m_{\rm L,c}$) and massive clusters ($M_{\rm clus}$).

\section{Self-lensing probability for von Zeipel-Lidov-Kozai mergers}
\label{appendix:ZLK}
Here we derive the optical depth for the von Zeipel-Lidov-Kozai induced BBH mergers that can occur in the potential of a massive BH \cite{petrovich-antonini-17}, following a similar approach as for the single-single GW captures in Appendix~\ref{appendix:gw-capture}. We derive simple scaling relations for the merger rate density, $\Gamma_{\rm ZLK}(r)$, that we compare to results of the Monte Carlo models of \cite{hoang-18}.

This mechanism operates on existing BBHs that formed from massive binary stars. This means that within a certain minimum distance to the central BH there are no binaries because they are all destroyed by the tides of the central BH.
This minimum distance depends on the BBH's semi-major axis $a$: $r_{\rm min}\sim a(m_{\rm L,c}/\MS)^{1/3}$ for circular orbits \cite{1988Natur.331..687H}, where as before $m_{\rm L,c}$ is the mass of the central BH (the lens).  
There is also a limit on the minimum $a$ of BBHs. Assuming that the binary is formed from stellar evolution, the initial stars need to be separated enough to evolve independently. 
This minimum $a$ depends on the masses of the progenitor stars, which for simplicity we assume here that scales as $a\propto \MS^{1/3}$ such that $r_{\rm min}$ is independent on $\MS$, so we can write
\begin{equation}
r_{\rm min}= r_6\,\left(\frac{m_{\rm L,c}}{10^6\,\msun}\right)^{1/3}, 
\end{equation}
where $r_6$ depends on details of stellar evolution and the eccentricity of both the BBH and its orbit around the central BH and we estimate its value  using the models of \cite{hoang-18}. In their Fig.~4, we see that $r_{\rm min}\simeq 100\,\au$ for $m_{\rm L,c}=10^7\,\msun$, such that $r_6\simeq50\,\au$. That figure also shows that $d\Gamma_{\rm ZLK}/d\log r$ in their ``Bahcall-Wolf-like" (BW) model is roughly constant, so we can write 
\begin{equation}
\Gamma_{\rm ZLK}(r) \equiv \Gamma_{\rm h}\left(\frac{r}{r_{\rm h}}\right)^{-3}.
\end{equation}
As in Appendix~\ref{appendix:gw-capture}, we determine the normalized probability for the radial distance of the source, $p_{\rm S}(r)$, as
\begin{equation}
p_{\rm S}(r) = \frac{\Gamma_{\rm ZLK}(r)}{\Gamma_{\rm tot}},
\end{equation}
where 
\begin{align}
\Gamma_{\rm tot} &= \int_{r_{\rm min}}^{r_{\rm max}}\Gamma_{\rm ZLK} 4\pi r^2dr,\\
&= 4\pi \Gamma_{\rm h}r_{\rm h}^3\ln\left(\frac{r_{\rm cusp}}{r_{\rm min}}\right),
\end{align}
where in the last step we used $r_{\rm max} = r_{\rm cusp}$ such that with the expression for $r_{\rm cusp}$ from Eq.~\eqref{eq:rcusprh} we have
\begin{align}
\frac{r_{\rm cusp}}{r_{\rm min}}&=\frac{r_{\rm h}}{10r_6}\left(\frac{m_{\rm L,c}}{10^6\,\msun}\right)^{2/3}\left(\frac{M_{\rm clus}}{10^7\,\msun}\right)^{-1},\\
&\simeq2\times10^3 ~,
\label{eq:rcusprmin}
\end{align}
where in the last step we adopted  $r_{\rm h}=5\,\pc$. We then find
\begin{equation}
p_{\rm S}(r)\simeq
(30\pi r^{3})^{-1}.
\end{equation}
The optical depth is then found as in Eq.~\eqref{eq:tau_app}
\begin{align}
\tau&\simeq\frac{R_{\rm L}\ymax^2}{15r_{\rm min}},\\
&\simeq3\times10^{-5}\left(\frac{m_{\rm L,c}}
{10^6\,\msun}\right)^{2/3} \ymax^2,
\end{align}
using $r_{\rm cusp}\gg r_{\rm min}$ (see Eq.~\eqref{eq:rcusprmin}).

\bibliography{arXiv_final}

\newpage

\end{document}